\documentclass[usenatbib]{mn2e}
\usepackage{graphicx}
\usepackage{epsfig}

\usepackage{amsmath}
\def\gsim{\mathrel{\raise0.35ex\hbox{$\scriptstyle >$}\kern-0.6em
\lower0.40ex\hbox{{$\scriptstyle \sim$}}}}
\def\lsim{\mathrel{\raise0.35ex\hbox{$\scriptstyle <$}\kern-0.6em
\lower0.40ex\hbox{{$\scriptstyle \sim$}}}}

\title[RX J1347-1145]{Large-scale Structure and Dynamics of the Most X-ray Luminous Galaxy Cluster Known --- RX J1347-1145}

\author[Lu et al.]{Ting Lu$^{1}$, David G. Gilbank$^{1}$, Michael L. Balogh$^{1}$, Martha Milkeraitis$^{3}$, Henk Hoekstra$^{2}$, 
\newauthor
 Ludovic Van Waerbeke$^{3}$, David A. Wake$^{4,5}$, Alastair C. Edge$^{4}$, Richard G. Bower$^{4}$\\ \\
$^{1}$Department of Physics and Astronomy, University of Waterloo, Waterloo, Ontario, N2L 3G1, Canada\\
$^{2}$Leiden Observatory, Leiden University, PO Box 9513, 2300 RA, Leiden, The Netherlands\\
$^{3}$Department of Physics and Astronomy, University of British Columbia, 6224 Agricultural Road, Vancouver, BC V6T 1Z1, Canada\\
$^{4}$Department of Physics, Durham University, South Road, Durham, DH1 3LE, UK\\
$^{5}$Department of Astronomy, Yale University, New Haven, CT, 06520, USA}

\date{\today}

\begin{document}
\maketitle

\begin{abstract}
We present photometric, spectroscopic and weak lensing analysis of the
large-scale structure and dynamics of the most X-ray luminous galaxy
cluster known, RX J1347-1145, at $z=0.451$. We spectroscopically confirmed 47
new members with LDSS3 on the Magellan telescope. Together with previously known  members, we measure a new
velocity dispersion of $1163\pm 97$ km s$^{-1}$. The mass inferred from
our velocity dispersion is $M_{200}= 1.16^{+0.32}_{-0.27} \times 10^{15}
M_{\odot}$, with  $r_{200}=1.85$ Mpc, under the assumption of a
singular isothermal sphere. We also present a weak lensing analysis
using deep CFHT data on this cluster, and find a deprojected mass of
$1.47^{+0.46}_{-0.43} \times 10^{15} M_{\odot}$ within $r_{200}$, in
excellent agreement with our dynamical estimate. Thus, our new
dynamical mass estimate is consistent with that from  weak lensing and
X-ray studies in the literature, resolving a previously claimed discrepancy. 
We photometrically detect and spectroscopically confirm another massive cluster with $\sigma=780\pm 100$ km s$^{-1}$ and  $M_{200}=3.4^{+1.4}_{-1.1} \times 10^{14} M_{\odot}$  $\sim 7$ Mpc south-west of RX J1347-1145, which we refer to as RXJ1347-SW. Our spectroscopic survey reveals a possible excess of galaxies in velocity space in the region between  RX J1347-1145 and RXJ1347-SW; comparing with simulations, this excess appears consistent with that expected from a large filamentary structure traced by galaxies connecting these two clusters.

\end{abstract}


\section{Introduction}

How galaxy properties are affected by the environment they reside in remains one of the unsolved puzzles of galaxy formation. The galaxy population in clusters has been observed to have different properties from that in the field \citep[e.g.][]{balogh04,w06b,mcgee08}, suggesting that this growth of large scale structure is accompanied by a transformation in galaxy properties.  However, it is still an open question as to exactly what mechanisms are responsible for this difference \cite[e.g.][]{B99,tanaka06}. 

Clues might come from the outskirt regions of clusters, because that is where field and groups of galaxies are accreting onto clusters, especially along filamentary structures that are seen in simulations \citep[e.g.][]{bond96,colberg99}. Groups in the outskirts are particularly interesting; a recent study by  \cite{mcgeeacc}, using a semi-analytic model, shows that a large fraction of galaxies that accrete onto clusters are in groups (but see \citealt{ber09}), and they might have been pre-processed in the group environment before they fall into clusters \citep{B00,fu04,li09,bm09}. However, these groups are likely to be different from isolated groups in the field, due to their earlier formation time \citep{mau} and the tidal fields from the large-scale structure around clusters \citep{tid}. Therefore, closer examination of groups and filamentary structures in the outskirts of clusters, and a comparison with isolated groups in the field, may provide important clues to how the cluster environment affects the properties of galaxies.   

We have started an extensive spectroscopic campaign on two contrasting clusters at intermediate redshift, focusing on groups in the outskirt regions out to $\sim 8$Mpc. The ultimate goal is to put strong constraints on the effect of the environments by comparing properties of groups in the infall regions with those of isolated groups, and with galaxy formation models. In this work, as the first paper of a series, we present results from a pilot study on the dynamics and large-scale structure of one of the clusters, RX J1347-1145 ($z$=0.451).

RX J1347-1145 is the most X-ray
luminous cluster in the ROSAT All-Sky Survey \citep[RASS, ][]{sch95}, and has been the subject of intense study, through spectroscopic \citep{ck}, X-ray   \citep{schindler97,allen02,allen08,ettori04,gitti04,gitti07,bradac08,ota08,mir08}, lensing \citep{fischer97,sahu98,kling,bradac05,bradac08,halkola,mir08,bro08}, and Sunyaev-Zel'dovich (SZ) effect \citep{point01,komatsu01,kita} analyses. The centre of the cluster was initially thought to be dynamically relaxed.
Although there are two very bright galaxies near the centre (usually referred to as the brightest cluster galaxy (BCG) and the second BCG), they are only 18 arcsec ($\sim$ 0.1Mpc) apart in the plane of sky, and $<100$ km/s in velocity. However, a discrepancy between the relatively low mass measured dynamically by \cite{ck}, compared with X-ray and weak lensing measurements, led to the suggestion that the cluster is actually undergoing a major merger.  This hypothesis has received support recently from the discovery of shocked gas in the centre of the cluster \citep{allen02,gitti04,ota08,bradac08,mir08,komatsu01}.

Most of the previous work is on the central region of the cluster; in this paper,  we focus on the dynamics and the large-scale structure of this cluster, using imaging data obtained with CFHT and spectroscopic data obtained with the Magellan telescope. We describe the data in \S \ref{dat}. In \S \ref{res}, we present the photometric and spectroscopic analysis on the large-scale structure, and the detection of another massive cluster, RXJ1347-SW, $\sim 7$ Mpc from the cluster centre. We also present mass estimates of the two clusters through dynamical methods and weak lensing analysis. In \S \ref{dis} we compare our mass estimates of the main cluster with those in the literature, and discuss the possible filamentary connection between the main cluster and  RXJ1347-SW using simulations. Our findings are summarized in \S \ref{sum}.

Throughout this paper, we assume a cosmology with $\Omega_m=0.3$, $\Omega_{\Lambda}=0.7$ and $H_o$=70 km s$^{-1}$ Mpc$^{-1}$. All magnitudes are in the AB system.

\section{Data}\label{dat}

\subsection{Imaging}

Deep $g',\ r'\ ,i\ ',z'$ band images of the cluster over a 1 square
degree field were obtained using Megaprime on the CFHT in semester 05A
(proposals 05AC10 and 05AC12), with total exposure times of 4200 s in
$g'$, 7200 s in $r'$, 1600 s in $i'$, and 3150 s in $z'$. The $z'$ band data will not be mentioned further, because they are not relevant for the work presented here.  In each
filter, images from single exposures are co-added using SWARP v2.16.0
\citep{swarp} and photometry is performed using SExtractor v2.4.4
\citep{sextractor}, with the camera run zero point. The latter quantity is provided by Elixir, and is extracted from the Landoldt field observed under photometric conditions every night. Since our science data were not taken under photometric conditions, a zero point offset calculated from  a short exposure taken under photometric conditions is applied. The typical seeing is $\sim 0.8''$, $\sim 0.7''$ and $\sim 1''$ in $g'$, $r'$ and $i'$ band respectively. Colours are measured  using magnitudes within a fixed aperture of diameter  4.7
arcsec, which is much larger than the seeing difference in
different filters. The magnitude measured within the Kron radius
($mag\_auto$) is considered as the total magnitude for each
object.  Although crowding and blending of objects affects the photometry for about ten percent of the galaxies in our sample, we have checked that none of the results presented here are sensitive to this.

Star-galaxy separation is based on the stellar locus defined
by the half-light radius as a function of total magnitude, in combination with the SExtractor parameter, $class\_star$. Objects with  $class\_star\leq 0.98$ in the $g'$ band, and not on the stellar locus, are considered  galaxies. The completeness limit for galaxies is  $\sim$0.8 mag brighter than the $5\sigma$ limit for  point sources \citep{yee91}; therefore, the completeness limit for galaxies in our catalog in the three bands that are relevant in this work is estimated to be  $g'\sim 25$ mag, $r'\sim 25$ mag, and $i' \sim$ 23.2 mag.

\subsection{Spectroscopy}
\subsubsection{Observations}\label{obs}
\begin{figure*}
\begin{center}
\includegraphics[width=18cm]{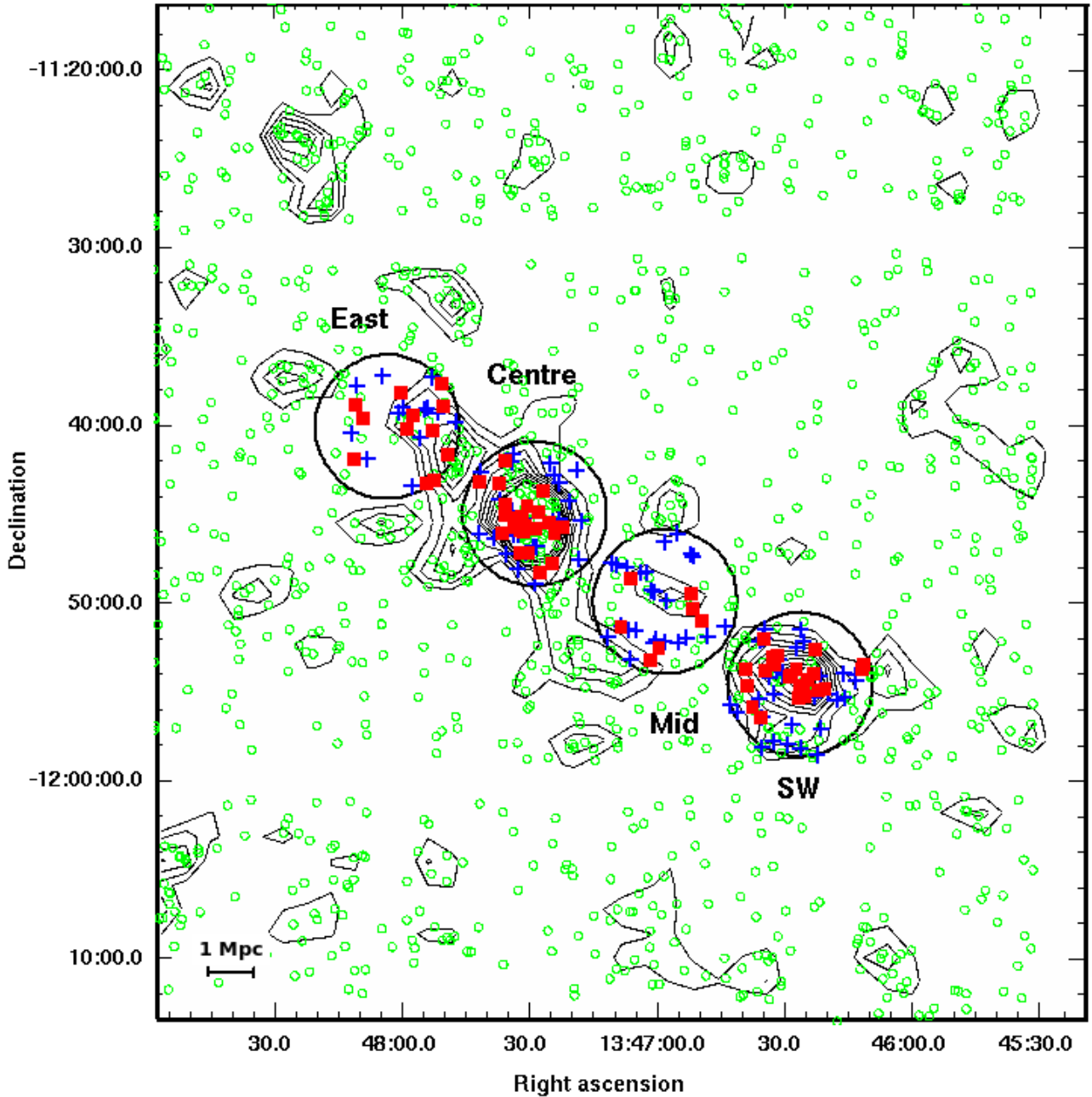}
\caption{The spatial distribution of the large-scale structure with the surface density contours of red galaxies overlaid. The density contours start at 2$\sigma$ from the mean and increase with an interval of 1$\sigma$. The small circles indicate all galaxies with red colours, the squares and crosses are respectively confirmed red and blue galaxies that are associated with the large-scale structure. The four large circles indicate the pointing positions with spectroscopic follow-up.\label{sig}}
\end{center}
\end{figure*}
In this section, we first describe our spectroscopic observation at the four pointings (shown in Figure \ref{sig}) that are chosen based on the large-scale structure traced by red-sequence galaxies, which will be described in  \S \ref{photst}.

To make the survey more efficient, we use the combination of $(g'-r')$ and $(r'-i')$ colour information to eliminate obvious higher redshift galaxies (see figure 6 in \citealt{lu09} for more details). Moreover, we select galaxies with $i'\leq 22$ mag. This ensures that for objects of interest here ($(g'-i')<3$) our sample is complete in each band, and the colour measurement is reliable. We make sure our targets are not colour biased; as can be seen in Appendix \ref{cp}, the completeness in each pointing for the red and blue populations separately is similar, at $\sim 50$ per cent.

Spectra were obtained using the Low Dispersion Survey Spectrograph 3 (LDSS3) on the 6.5 meter Clay Telescope over the nights from 2008 April 28th to May 1st. The pointing positions are indicated in Figure \ref{sig}, each with a FOV of 8.3 arcmin in diameter (corresponding to r=$1.4$ Mpc at the redshift of the cluster). The VR filter and Medium Red grism were used, which covers the  wavelength range 5000\AA \  to 7200\AA . At the redshift of the cluster, the important spectral features CaH, CaK, G-band, H$\delta$ and [OII]  fall into this wavelength range. Each pointing was observed with two masks, with an average of $\sim 60$ objects and 3x20 mins exposure, per mask, with slit width 1.2 arcsec, providing a spectral resolution of $\sim 14$\AA.

\subsubsection{Data Reduction}
 The 2d raw spectra were first reduced using the COSMOS2 software package, then 1d spectra were extracted and redshifts were determined using a modified version of $zspec$ kindly provided by Renbin Yan \citep{deep2}. In $zspec$,  spectra are first compared with eigen-templates and the ten best-fitted redshifts are provided. We then visually examine the spectra and determine the best redshift. Usually the first or second choice provided is a good fit. Occasionally when none of the ten choices is a good match, we manually cross-correlate the spectral features and find the best-fit redshift.  We consider redshifts determined from at least three supporting spectral features as secure. Two example spectra, near the magnitude limit of our survey, are shown in Appendix \ref{eg}. 

For the 473 objects targeted, we obtained 400 secure redshifts, which are used in the analysis here. 
Further details on the success rate, and the efficiency of our survey are presented in Appendix \ref{cp}.  From the objects that have duplicate observations, the uncertainty on the redshift is estimated to be $\sim 113$ km s$^{-1}$ in the rest frame (see Appendix \ref{zun}). We also compare the redshift of the common objects in our catalogue with that from \cite{ck}, and find no systematic difference between these measurements (Appendix \ref{zun}); therefore, in the following analysis, we add the extra 34 redshifts from \cite{ck} to our sample when applicable.

\section{Results}\label{res}

\subsection{Photometrically Identified Large-Scale Structure}\label{photst}

In this section, we describe the large-scale structure traced by the red-sequence galaxies. Again, only galaxies with $i'\leq 22$ mag are used, for reasons discussed in \S \ref{obs}. At the redshift of this cluster ($z=0.45$), the $(g'-i')$ colour brackets the 4000\AA \ break; therefore we use this colour to identify red-sequence galaxies around this cluster, using a method based on the Red-sequence Cluster Survey technique  \citep[e.g.][]{GY00,lu09}. We define the red-sequence at the redshift of the cluster using known cluster members from \cite{ck} that have $(g'-i')$ colour measurement in our catalogue. Given the small slope and the  scatter of the colour, we choose to use a red-sequence that has zero slope with a half width of 0.15 mag. Assuming the probability distribution of the colour of each galaxy is
Gaussian, with $\sigma$ being the error on the colour measured from SExtractor, we can calculate the probability of each galaxy belonging to
the colour slice (for details see \citealt{GY00}). The fact that the colour error from SExtractor is likely to be underestimated is not important here, because the width of the colour slice used is much wider than the colour uncertainties.  We consider galaxies with at least 10 per cent probability as red-sequence galaxies, and they are plotted as small circles
in Figure \ref{sig}. Our results here are insensitive to the
details of how the red-sequence galaxies are selected.

Using the background number counts measured from
the  Canada-France-Hawaii Telescope Legacy Survey (CFHTLS) Wide field
1, we can calculate the overdensity of red-sequence galaxies. We define
the significance of the overdensity as the background-subtracted number
of red-sequence galaxies  divided by the r.m.s. of the background
distribution. The contours of the significance map, smoothed over a
region of radius $\sim 0.5$ Mpc with a top hat function, are plotted in
Figure \ref{sig}, starting from 2$\sigma$, with an interval of
1$\sigma$.  

The most significant peak ($\sim 25\sigma$) is the main cluster (the pointing ''Centre''). The number of red-sequence galaxies (background subtracted) within 0.5 Mpc from the cluster centre, brighter than $m^*+2$ ($i'=22$ mag), denoted  $N_{red,m^*+2}$, is $\sim 46\pm 7$, where the uncertainty is estimated using Poisson statistics. Converting $N_{red,m^*+2}$  to $N_{200}$ according to figure 8 in \cite{lu09}, and using the scaling relation between $N_{200}$ and  $M_{200}$ \citep{johnston,hansen07}, we estimate the mass of the main cluster from its richness to be $M_{200}\sim 1.4^{+0.4}_{-0.3}\times 10^{15} M_{\odot}$; although we note the conversion from  $N_{red,m^*+2}$ to $N_{200}$ for such a high richness system is extrapolated from lower richness systems. 

There is another prominent peak with a significance of $\sim 10\sigma$, about 7 Mpc (projected) away from the main cluster (pointing ``SW'').  The $N_{red,m^*+2}$ for this system is $\sim 22\pm 5$, which corresponds to $M_{200}=(3.3^{+1.3}_{-1.6})\times 10^{14} M_{\odot}$. This suggests this system is a cluster in its own right, and we will refer it as RXJ1347-SW.

The significance contours also reveal several clumps of overdensities in the infall region around the main cluster, and an elongated large-scale structure running from NE to SW, along a line connecting the main cluster and RXJ1347-SW. Four positions along this line are targeted for our spectroscopic follow-up with the Magellan telescope as described in \S \ref{obs}. (Note the significant overdensity east of the pointing ``Mid'' has been targeted with our on-going Gemini program.)

\subsection{Dynamics}
\begin{figure}
\leavevmode \epsfysize=8cm \epsfbox{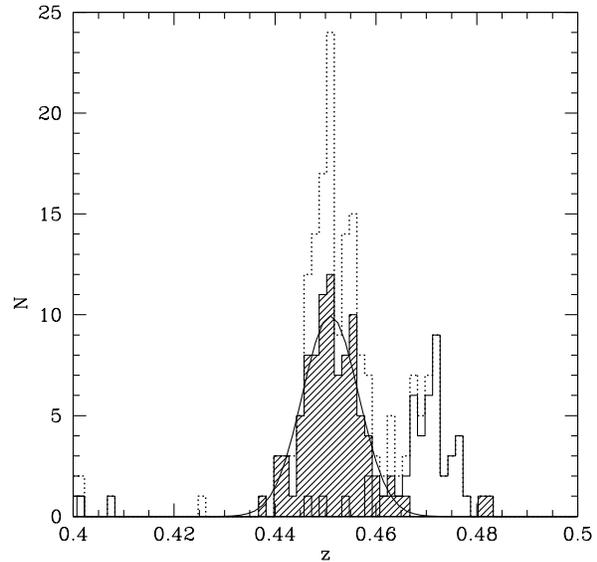}
\caption{The galaxy redshift distribution from the full area is shown as the dotted, open histogram. The shaded histogram shows the distribution of galaxies in the centre of the main cluster ($r\lsim 1.4$ Mpc), with a Gaussian function that is statistically consistent with this distribution, overlaid. The redshift distribution of galaxies within $r\lsim 1.4$ Mpc around RXJ1347-SW is shown as the solid open histogram.\label{z}}
\end{figure}

In this section, we present the dynamical analysis of the above described photometrically identified structure, based on the redshifts obtained.
\subsubsection{Redshift Distribution and Velocity Dispersion}\label{vdisp}
 In Figure \ref{z}, the full redshift distribution of all galaxies is shown as the dotted open histogram. There are two main peaks in the full distribution: one corresponds to the centre of the main cluster, and the other corresponds to RXJ1347-SW. We analyze these two clusters in detail first in this section, and examine the dynamics of the large-scale structure in the next section. 

 The redshift distribution of galaxies within $\sim 1.4$ Mpc from the main cluster centre is shown as the shaded histogram. The distribution is smooth in general; there is an indication of two separate peaks, but the distribution is still consistent with being drawn from the single Gaussian overplotted on the shaded histogram in Figure \ref{z}. In particular, a $\chi^2$ test shows that a subsample drawn from this single Gaussian distribution would have a $\chi^2$ value as large as we observed
 about $80$ per cent of the time.  
 The mean redshift we measure is $0.4513\pm0.0006$, consistent with the previous measurement by \cite{ck}. The dispersion we measure, using the bi-weight method of \cite{bi-wt}, with 3$\sigma$ clipping applied, is $0.0056\pm 0.0005$, which corresponds to a rest-frame velocity dispersion of 1163$\pm 97$ km s$^{-1}$. The errors are estimated using Jackknife resampling \citep{jk}. From now on, we refer to galaxies that are within 3 times the dispersion as members of the main cluster. The velocity dispersion we measure is higher than that of \cite{ck} (we discuss this more in \S \ref{lit}).

Despite the consistency with a single Gaussian there is an indication of two separate velocity peaks in the distribution.  A $\chi^2$ test shows that, at the $\sim$30 per cent level, the distribution is consistent with a combination of two Gaussian functions: one with $\bar{z_1}=0.4495$ and rest-frame velocity dispersion $\sigma_1=620$ km/s, and the other with $\bar{z_2}=0.4552$ and $\sigma_2=535$ km/s. In that case, the centre of the cluster would be composed of two separate components with much smaller masses, and a relative velocity of $\sim 1000$ km s$^{-1}$; considerably larger than the $<100$km/s difference between the two BCGs ($z_1=0.4515$ and $z_2=0.4511$).   Although statistically this model is not favoured over the single-population model, we cannot rule out the possibility that there are two distinct dynamical components.  We revisit this further in \S~\ref{mer} when we consider both the spatial and velocity distribution of galaxies.

The redshift distribution of RXJ1347-SW (pointing ``SW'') is shown as the open solid histogram in Figure \ref{z}. It is not significantly different from a Gaussian distribution, with a mean redshift of $0.4708\pm 0.0006$, and dispersion of $0.0038\pm 0.0005$, or $780\pm100$ km s$^{-1}$.

\begin{figure}
\leavevmode \epsfysize=8cm \epsfbox{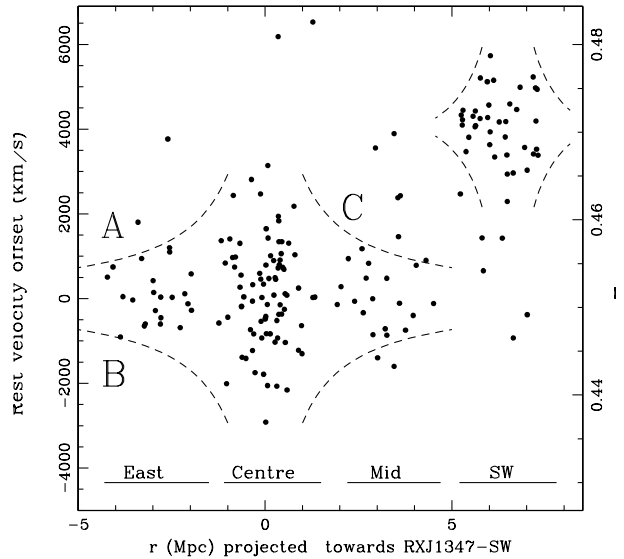}
\caption{The rest frame velocity offset as a function of distance relative to the cluster centre projected along the line towards RXJ1347-SW. The dashed curves indicate the boundary of the caustics. See text for details.\label{vr}}
\end{figure}

\subsubsection{Phase Space Distribution}\label{voffset}

We examine the dynamics of the large-scale structure by  converting the
redshift of each galaxy to a rest-frame velocity relative to the
central redshift of the main cluster. In Figure \ref{vr} we show the
velocity offset as a function of position projected along the line
connecting the main cluster and RXJ1347-SW. Galaxies in the inner
region of the cluster mainly occupy regions spanning from $-2000$ km
s$^{-1}$ to 2000 km s$^{-1}$, symmetrically, while RXJ1347-SW is at
$\sim 4000$ km s$^{-1}$ from the main cluster. Note that in velocity
space, the region occupied by members of massive clusters resembles a
trumpet shape, as expected in a spherical infall model \citep[e.g.][]{har93}. The dashed curves in Figure \ref{vr} indicate
the boundary of the caustics for these two systems (as will be discussed in \S \ref{mdyn}), derived analytically from equation (7) in \cite{d99} assuming 
the two systems follow a NFW profile \citep{NFW}. In pointing ``East'', on the other side of the main
cluster away from RXJ1347-SW, there are almost no galaxies in the
regions labelled ``A'' and ``B'' in Figure \ref{vr} around the cluster
above the caustic curve.  In comparison, there are $\sim 8$
galaxies occupying the region between the main cluster and RXJ1347-SW,
labelled ``C''.  Given that our redshift completeness is about 50 per
cent, we expect there are approximately twice as many $i'<22$ galaxies
($\sim 16$) within that region in total. Note this is not a sampling bias, because the pointing ``East'' and ``Mid'' are located symmetrically on two sides of the main cluster. Also, the uncertainties on the redshift determination, or the position of the caustic curve, would not change the fact that there is an excess of galaxies between the main cluster and RXJ1347-SW. In \S \ref{simu} we will use simulations to investigate whether these galaxies are physically residing in regions between the main cluster and RXJ1347-SW, and thus suggest a possible structure connecting the two systems.

\subsection{Colour Distribution}\label{color}

As described in \S \ref{photst}, we trace the large-scale structure around this cluster using red-sequence galaxies. We consider galaxies in the redshift range $\bar {z}_{cl}-3\sigma_{cl}<z< \bar {z}_{sw}+3\sigma_{sw}$ as associated with the structure, where $\bar {z}_{cl}$ and $\sigma_{cl}$ are the mean and dispersion of the redshift distribution of the main cluster, and $\bar {z}_{sw}$ and $\sigma_{sw}$ are the same for RXJ1347-SW. In Figure \ref{sig}, solid squares and crosses represent red and blue galaxies  that are spectroscopically confirmed to be associated with the structure, respectively. 

\begin{figure}
\leavevmode \epsfysize=8cm \epsfbox{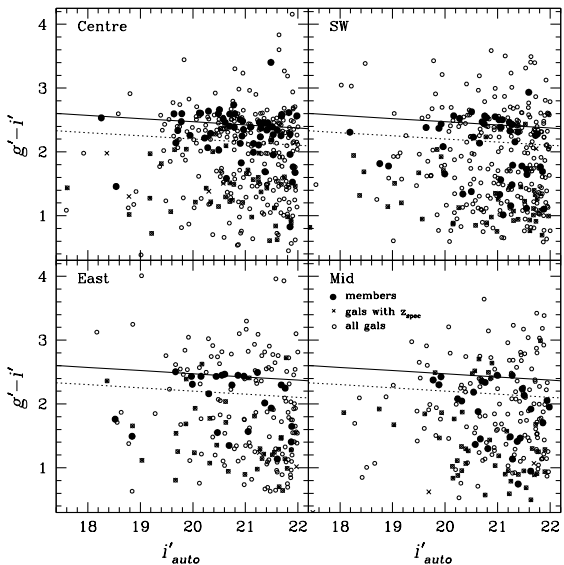}
\caption{The colour-magnitude diagram of galaxies is shown for each of the four pointings we targeted, along the elongated structure. Open circles are all galaxies in the field of view, crosses are galaxies with secure redshifts, and filled circles are members associated with the structure. Solid lines are fitted red-sequences, and dashed lines separate blue galaxies from the red-sequence (see text for detail).\label{cmd}}
\end{figure}

The study by \cite{BO} revealed a higher fraction of blue galaxies in high redshift clusters compared to local ones (the Butcher-Oemler effect), indicating on-going star formation in high redshift clusters. Since then there have been many studies on the dependence of this effect on the environment, and cluster properties such as the X-ray temperature \citep[e.g.][]{ur09,wake05}. RX J1347-1145 is the most X-ray luminous cluster known, and has an elongated large-scale structure; therefore it is interesting to explore the colour distribution of the galaxies in this structure. The colour-magnitude diagram of galaxies in the four pointings along the structure is shown in Figure \ref{cmd}. Open circles are all galaxies in the field of view, crosses are galaxies with secure redshift obtained, and filled circles are members associated with the structure.

We calculate the blue fraction of the main cluster using the
\citet{BO} criterion, i.e. galaxies 0.2 mag bluer than
the CMR in rest-frame (B-V). We transform rest-frame $\triangle
(B-V)=0.2$ into our observed $\triangle(g'-i')$ using the colour
difference between E and Sab galaxies from \cite{fuku}, as indicated by the dashed lines in Figure \ref{cmd}. The  number of
blue cluster members is calculated by dividing the number of
spectroscopically confirmed blue galaxies by the completeness of the
spectroscopy for the blue population. The number of red cluster members
is calculated in the same way, and the total number of members is then
the sum of that of the blue and red galaxies. The blue fraction of RX
J1347-1145  within the central 1.4 Mpc ($\sim 0.8 r_{200}$), down to
$i'=21$ (corresponding to $M_{V}\sim -20$) is  $20_{-8}^{+9}\%$. The
uncertainty is estimated using the weighted binomial statistics \cite{bino}.  Note although RX J1347-1145 is the
most X-ray luminous cluster in the RASS, its blue fraction in the
central region is similar to that of other less luminous clusters \citep{ellingson01}, consistent  with the lack of X-ray luminosity dependence found by  \cite{wake05}. The blue fraction of the cluster RXJ1347-SW ($36_{-12}^{+13}\%$) is, within the uncertainty, consistent with that of the main cluster.

From the contours in Figure \ref{sig},  the region between the main cluster and RXJ1347-SW shows an overdensity of red-sequence galaxies. Our spectroscopy also reveals a large population of blue galaxies, and it is interesting to note that they form two trails running in the same direction as the elongation of the large-scale structure. The large blue population associated with an overdensity of red galaxies is particularly interesting. It suggests that targeting regions with overdensities of red-sequence galaxies for spectroscopic follow-up is a strategy that allows the efficient selection of cluster members, while still tracing interesting structures with active star formation. A more thorough analysis of the stellar populations in the large-scale structure will be done when the rest of the data are obtained.

\subsection{Mass Estimate}\label{massest}
In this section, we estimate the mass of the main cluster and RXJ1347-SW using different methods. We defer a comparison with the literature, and the discussion of the dynamical implication, to \S \ref{dyn}. 

\begin{figure*}
\includegraphics[width=8cm]{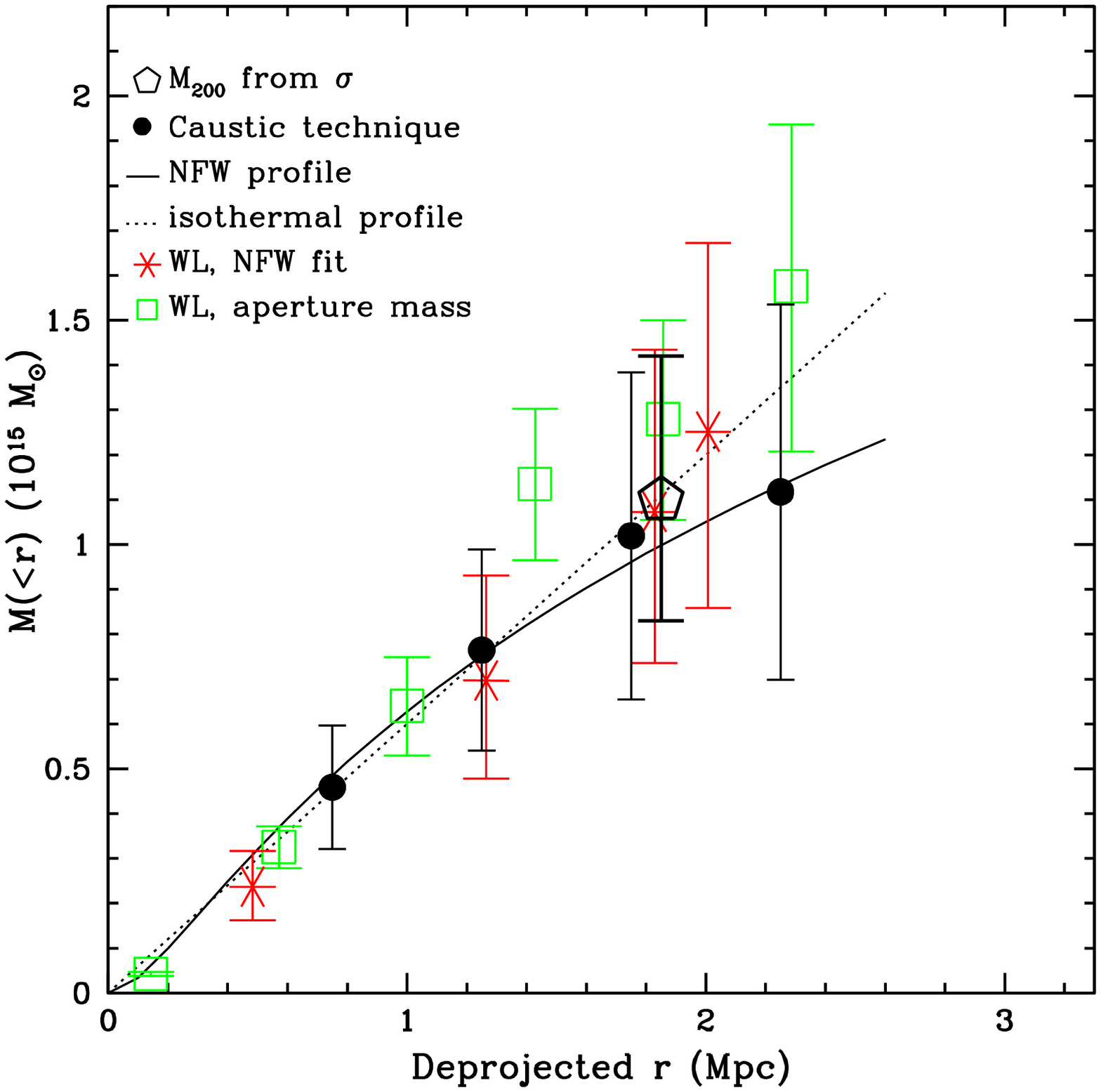}\includegraphics[width=8cm]{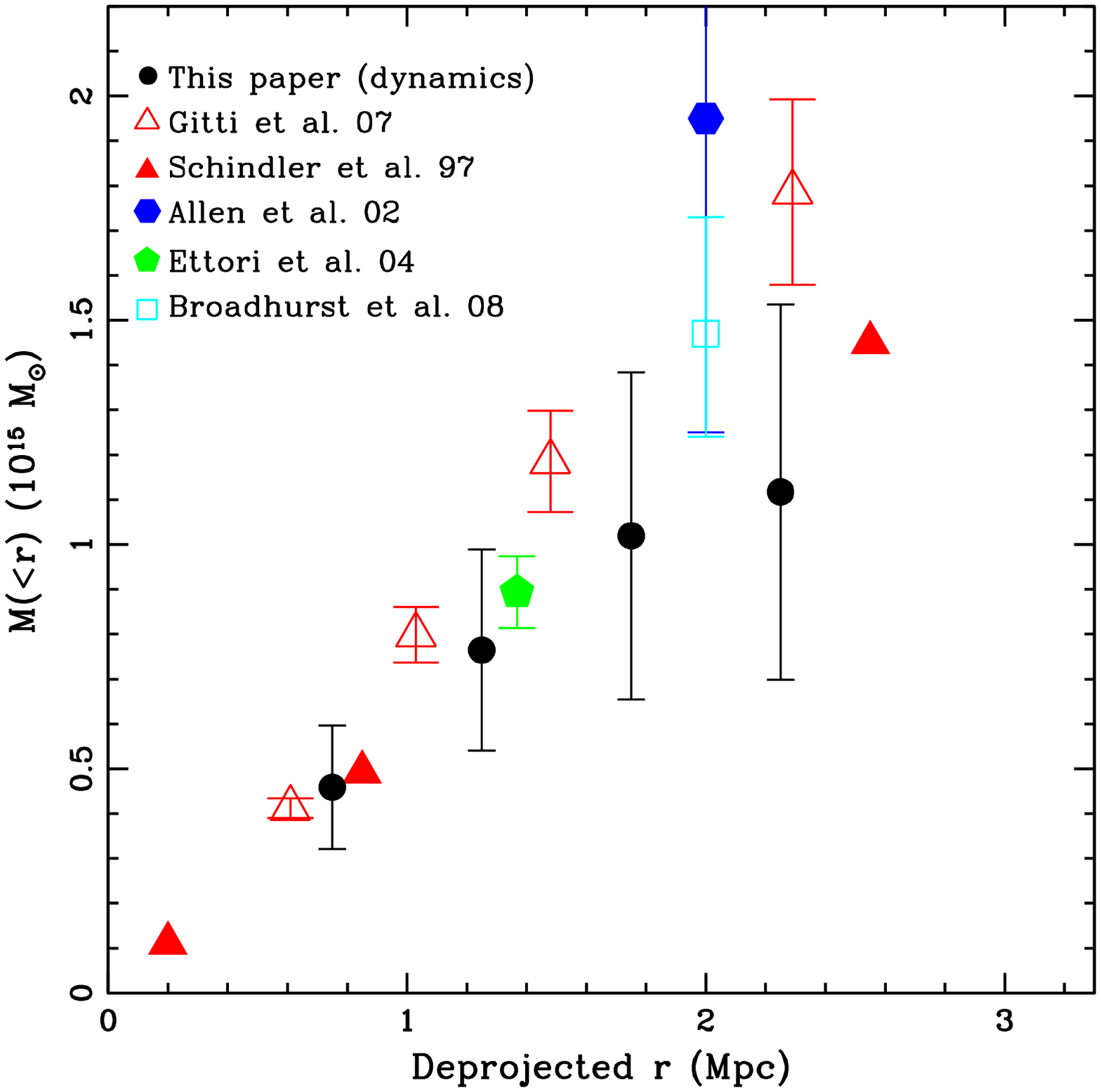}
\caption{The deprojected mass profile for RX J1347-1145. Left panel: the black points  represent the mass profile inferred from the caustic method. The solid, and dotted curves show the fitted NFW and isothermal profile respectively. The stars and open squares are our weak lensing analysis. Right panel: the black points are the same as in the left panel. The other symbols are mass estimates from the literature.
\label{mcau}}
\end{figure*}

\subsubsection{Dynamical Estimates}\label{mdyn}

For a given density profile and velocity anisotropy, the velocity dispersion can be used to infer the virial mass. Under the assumption of a singular isothermal density profile and a velocity anisotropy that satisfies $\sigma=v_c/\sqrt{2}$ \citep{lokas01}, the virial mass can be expressed as $M_v=2\sigma_1^2 r_v/G$. Applying this to the measured velocity dispersion (\S \ref{vdisp}) yields a $M_{200}=\sqrt{2}\sigma_1^3/\left[{5H(z)G}\right]=1.16^{+0.32}_{-0.27} \times 10^{15} M_{\odot}$, with $r_{200}=\sqrt{2}\sigma/\left[{10H(z)}\right]=1.85$ Mpc.  This mass estimate agrees well with that inferred from the richness of the cluster, as described in \S \ref{photst}.

Under the same assumptions, the mass of RXJ1347-SW is estimated to be $M_{200}= 3.4^{+1.4}_{-1.1} \times 10^{14} M_{\odot}$, with $r_{200}=1.22$ Mpc. This $M_{200}$ is also in good agreement with that inferred from the richness, and confirms that the cluster RXJ1347-SW we detected photometrically using the red-sequence galaxies is indeed a massive cluster. There has not been any report on the existence of this cluster in the literature.

 Another method to estimate the mass through dynamical analysis is the caustic technique, under the assumption that  clusters form through hierarchical accretion \citep{d97,d99}. The advantage of this technique is that it does not depend on the assumption of dynamical equilibrium, and it can probe the mass out to large radii where the virial theorem does not apply any more.

As described in \S \ref{voffset} (Figure \ref{vr}),  cluster members occupy a region in velocity space with a trumpet shape delineated by caustics that separate them from the Hubble flow. Assuming spherical symmetry, the mass of the cluster can be estimated as:
\begin{eqnarray}
M(<r)=\frac{1}{2G}\int_o^r \mathcal{A}^2(r)dr,
\end{eqnarray}
where  $\mathcal{A}$ is the amplitude of the caustic. The prescription for implementing the caustic method is described in detail in \cite{d99}. Here we only summarize the key steps. Since the location of the caustics is where clusters are separated from the Hubble flow, it is indicated by a significant drop of the density of cluster members in velocity space. Therefore, one first calculates the density as a function of velocity at different radii, using an adaptive kernel method (Silverman 1986). In order to use the spherical smoothing window, the window size in velocity and distance, $h_v$ and $h_r$, has to be rescaled \citep[see ][for details]{d99}. For our data, we choose $q=h_v/h_r=6$, and a spherical smoothing window size of $\sim$ 700 km/s\footnote{Within the range of reasonable choices of parameters, the change of the resulting mass profile is within the resulting error bars.}.  In the central region of clusters, the escape velocity (essentially the location of the caustics) is twice the velocity dispersion, which is valid under  the assumption that the central region of the cluster is virialized. Therefore, in practice, the threshold  density where the significant drop occurs (and thus determines the amplitude of the caustics) is chosen so that the above condition is satisfied.

The deprojected mass profile of the main cluster estimated using this technique is shown as solid dots in both panels of Figure \ref{mcau}. The errors are estimated using Jackknife resampling \citep{jk}. At $r_{200}=1.85$ Mpc  
 the enclosed mass is $\sim 1 \times 10^{15} M_{\odot}$. This is in excellent agreement with that inferred from the velocity dispersion; however, this is largely true by construction, since the density threshold is chosen such that the two techniques agree within the virialized region of the cluster.  The caustic-based measurements have the most value at larger radii, where the assumption of virial equilibrium is unlikely to hold.  Note beyond $r\sim2$ Mpc, we only have sparse sampling in two pointings, and the velocity distribution of galaxies in these pointings is not likely representative of that of the whole population at that radius, due to the presence of elongated large-scale structure. 
We fit the mass profile to a NFW \citep{NFW} and isothermal profile, shown as the solid and dotted curves respectively in the left panel of Figure  \ref{mcau}\footnote{The fitted NFW profile is used in the analytic derivation of the caustic curves shown in Figure \ref{vr} and discussed in \S \ref{voffset}.}. Because of the large uncertainties on the mass profile due to limited data, the two profiles cannot be distinguished.

\subsubsection{Weak Lensing Analysis}\label{len}
We use the deep CFHT $r'$ imaging data to study the cluster mass
distribution using weak gravitational lensing. Unlike dynamical
methods, the weak lensing signal provides a direct measurement of the
{\it projected} mass, irrespective of the dynamical state of the
cluster.  Furthermore, the observed signal can be used to ``map'' the
distribution of dark matter \citep[e.g.][]{ks93}.

Our cluster weak lensing analysis follows in detail the steps
described in \cite{henk07}. In order to relate the lensing signal to
a physical mass requires knowledge of the redshift distribution of the
sources. The wavelength coverage and limited depth of the observations
prevent us from deriving accurate photometric redshifts for the
individual sources. Instead we use the results from \cite{ilbert06}  who determined photometric redshifts for galaxies in the CFHTLS deep fields.

To determine the lensing signal, we measure the shapes of galaxies
with $22<r'<25$.  We use such a relatively faint cut because of the
high redshift of the cluster. The galaxy shapes are corrected using
the method presented in \cite{kaiser95}, with modifications described
in \cite{henk98}. The weak lensing signal is sensitive to all
structure along the line-of-sight and, as shown in
\cite{henk01,henk03}, this distant (uncorrelated) large-scale
structure introduces additional noise to the mass estimate. This
`cosmic shear' noise is included in the errors listed below. We note
that the noise due to the intrinsic ellipticities of the sources
dominates because of the relatively high cluster redshift.

\begin{figure}
\leavevmode \epsfysize=8.5cm \epsfbox{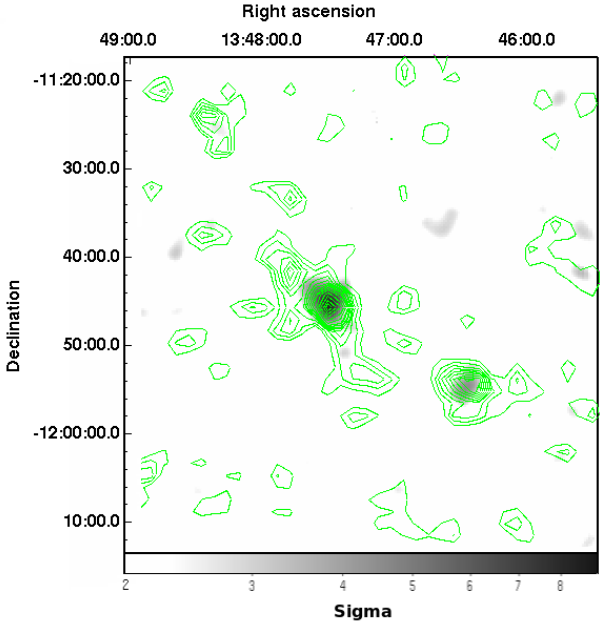}
\caption{The grey scale shows the weak-lensing signal-to-noise map, out to $\sim 10$ Mpc from the main cluster centre. Contours of the red galaxy surface density (as in Figure \ref{sig}) are overlaid, starting from 2$\sigma$ with an interval of 1$\sigma$. The two prominent peaks are the main cluster and RXJ1347-SW; their mass concentrations coincide with the overdensities of the red-sequence galaxies.
\label{s2n}}
\end{figure}

The weak lensing reconstruction of the projected surface density with
a smoothing scale of 123.3$''$ is presented in Figure~\ref{s2n}\footnote{The map presented in this Figure was constructed using source galaxies with $23<r'<25.5$} out to
$\sim 10$ Mpc. To indicate more clearly which structures are
significant, we show the signal-to-noise ratio, rather than the signal
itself, as the grey scale (see the left panel in Figure \ref{dres} for a more detailed
view). The main cluster and
RXJ1347-SW are clearly detected, at $\sim 8 \sigma$ and $\sim 4
\sigma$ levels respectively. No other significant structures are
evident in this map. The contours of the overdensity of the
red-sequence galaxies, the same as in Figure \ref{sig}, are
overplotted on top of the mass map. For the main cluster and
RXJ1347-SW, the mass concentrations coincide with the overdensities of
red-sequence galaxies.

\begin{figure}
\leavevmode \epsfysize=8cm \epsfbox{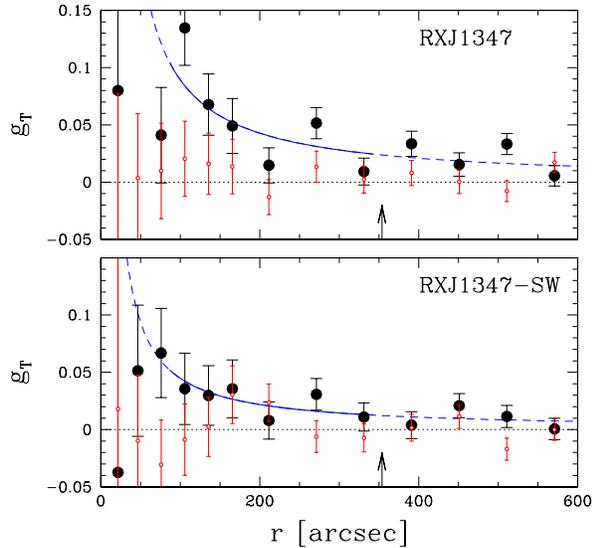}
\caption{Tangential distortion as a function of the distance from the cluster
center for the main cluster (top panel) and RXJ1347-SW (bottom panel).
In both cases we detect a significant lensing signal out to large
radii (the arrows indicate 2~Mpc). The open points correspond to the signal
that is measured when the sources are rotated by 45 degrees, which provides
a  measure of the level of systematics. The points are consistent with
zero. The blue dashed line shows the best fit SIS model, where the solid
blue part indicates the region to which the model was fit. The errorbars indicate the statistical error in the measurement, due to the intrinsic ellipticities
of the sources (see \protect \citealt{lenerr} for details). 
\label{gtprof}}
\end{figure}

The smoothing of the mass map limits its use to determine the cluster
mass. Instead, we compute the mean tangential distortions as a function
of distance from the cluster center. The resulting measurements for the
main cluster and RXJ1347-SW are shown in Figure~\ref{gtprof}.
We estimate the cluster masses from these shear measurements 
using two different methods. The first,
and most commonly used, is to assume a density profile and determine
the best fit parameters to the observed tangential distortion. For
comparison with the dynamical results, we first fit a singular
isothermal sphere model (SIS) to the observations between 500 kpc and
2 Mpc, which yields a velocity dispersion of $\sigma=1282\pm124$ km/s,
in excellent agreement with the results based on the redshifts of the
cluster galaxies. Thanks to the wide-field lensing data, we can also
apply the same method to  RXJ1347-SW (centred at 13$^{\rm h}$46$^{\rm
  s}$27.2$^{\rm s}$, $-11^\circ$54$'20.2''$), for which we obtain a
velocity dispersion of $\sigma=927\pm165$ km/s, also in good
agreement with the value inferred from the redshifts of the cluster
members.

Although the SIS allows for an easy comparison to the observed
velocity dispersion, a more physically motivated function is the NFW
\citep{NFW} profile, which has been inferred from numerical
simulations. We assume the mass-concentration relation from
\cite{duffy08} and fit the NFW model to the data within 500 kpc and
2Mpc. With these assumptions, we find a best-fit value for the virial
mass of $M_{\rm vir}=1.89^{+0.59}_{-0.55}\times 10^{15}M_\odot$. The
best fit NFW model within different apertures are indicated by stars
in the left panel of  Figure~\ref{mcau}. For RXJ1347-SW we obtain $M_{\rm
  vir}=6.4^{+3.9}_{-3.3}\times 10^{14}M_\odot$.

To compare to the results from the cluster galaxy dynamics, we also
compute the corresponding masses within $r_{200}^{\rm
  dyn}=1.85$~Mpc. For this choice of radius we obtain
$M_{200}=1.47^{+0.46}_{-0.43}\times 10^{15}M_\odot$ for the main
cluster. This result is in excellent agreement with the value inferred
from dynamics. For RXJ1347-SW we obtain $M_{200}=4.8^{+2.9}_{-2.5}\times
10^{14}M_\odot$ for $r_{200}^{\rm dyn}=1.22$~Mpc.

\begin{figure}
\leavevmode \epsfysize=8cm \epsfbox{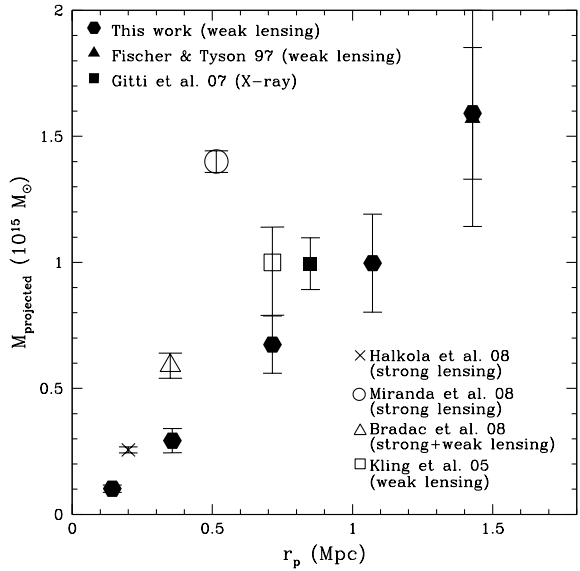}
\caption{The projected mass as a function of radius for RX J1347-1145. The solid hexagons are estimates from this work, and the other symbols are from the literature. See text for details.\label{m_proj}}
\end{figure}

Since weak lensing signals are sensitive to the total mass projected along the line-of-sight; it is interesting to consider a non-parametric estimate for
the projected mass within an aperture (e.g. \citealt{clowe98,fahlman94}; see Hoekstra
  (2007) for our implementation).  The resulting
{\it projected} masses as a function of radius are presented in
Figure~\ref{m_proj} as solid hexagons. Figure~\ref{m_proj} also shows
other estimates from the literature, which will be discussed in detail
in \S\ref{lit}.

As discussed in Hoekstra (2007) and \cite{mahdavi08} one can assume
spherical symmetry to deproject the resulting masses. We assume an NFW
profile to extrapolate along the line-of-sight, but note that the
results at large radii do not depend strongly on this assumption.  The
resulting deprojected masses are shown as open squares in the left panel of Figure
\ref{mcau}. They are consistent with our dynamical mass estimate using
the velocity caustics, within the uncertainties. 

Within an aperture of $r_{200}^{\rm dyn}=1.85$~Mpc, we find a
deprojected mass for the main cluster of $M_{200}=1.25\pm0.19\times
10^{15}M_\odot$, again in excellent agreement with the dynamical
measurements. Similarly we find a mass of $M_{200}=5.6\pm1.6\times 10^{14}
M_\odot$ for RXJ1347-SW using $r_{200}^{\rm dyn}=1.22$~Mpc, which is somewhat
larger than the dynamical mass.

\section{Discussion}\label{dis}
\subsection{Dynamical State of the Main Cluster}\label{dyn}
The mass estimate and the dynamical state of this cluster have been the focus of many previous studies. In this section, we first compare the mass estimated using different methods in this work and in the literature, and then discuss the dynamical state of the cluster.

\subsubsection{Comparison of Mass Estimates}\label{lit}

The mass of  RX J1347-1145 has been estimated using different techniques. Other than through the dynamics and weak lensing signals as described in \S \ref{massest}, X-ray data have also been used to infer the mass. 

In a recent study by \cite{gitti07}, they
estimate the mass by finding a NFW mass profile \citep{NFW} that
reproduces the observed X-ray temperature profile. The resulting
integrated mass at different radii are plotted as open triangles in the right panel of 
Figure \ref{mcau}. This is consistent with our estimates from both the
dynamics and the weak lensing analysis (\S \ref{massest}).

 Similarly, the study by \cite{allen02} estimated the virial mass by fitting a NFW profile to their Chandra observations, and their $M_{200}$ is plotted as the solid hexagon in the right panel of Figure \ref{mcau}. Other mass estimates using X-ray data by \cite{schindler97} and \cite{ettori04} are plotted in the right panel of Figure \ref{mcau} as well. All these estimates at different radii are consistent within their error bars.

As discussed in \S \ref{len}, the projected mass is a more direct measurement than the {\it{deprojected}} mass from the weak lensing analysis, since it is the total projected mass along the line-of-sight that gives rise to the weak lensing signals; therefore, it is more direct to compare the measured projected mass from different studies. 

In Figure \ref{m_proj}, the projected mass enclosed within a certain
radius estimated from strong, weak, and combined lensing analyses in
the literature are plotted, along with our weak lensing estimate (see
legends). Our weak lensing mass estimate (solid hexagons) is consistent
with that of \cite{fischer97} (the solid triangle), and these two estimates
are themselves in agreement with the X-ray analysis of \cite{gitti07} (the solid
square). 

The weak lensing mass estimate by \cite{kling} is
slightly higher ($\sim 2\sigma$). This discrepancy is likely due to differences in the assumed source
redshift distribution; \cite{kling} base their estimate on $r<24.5$
spectroscopy over the small area of the Hubble Deep Field, while we use
the photometric redshift distribution from the CFHTLS Deep survey.
Given our good agreement with both X-ray and dynamical mass estimates,
it seems likely that their analysis underestimates the mean source
redshift, and thus overestimates the cluster mass. 

 In comparison, the mass estimates from strong lensing studies \citep{halkola,mir08} and the combined strong and weak lensing analysis by \cite{bradac08} are systematically higher\footnote{We did not include the work by \cite{sahu98} because the redshift of the arc they assumed is not correct.}.  In general strong lensing provides an accurate estimate of the projected mass only within the central few hundred kpc,  whereas the weak lensing mass is better constrained at larger radii; therefore caution should be taken when  combining strong and weak lensing analysis. See also \cite{halkola}
 for a summary and more discussion of the comparison among these
 works.

 The good agreement shown in Figures \ref{mcau}
 and \ref{m_proj} suggest that the mass estimated from dynamical,
 X-ray, and weak lensing studies is generally consistent,  while strong
 lensing gives a slightly higher value.

So far, there has been only one other published redshift survey on this cluster, by \cite{ck}. They derived from 47 cluster members in their sample a velocity dispersion of 910$\pm 130$ km s$^{-1}$, using the same bi-weight method by \cite{bi-wt} as used in this work. This is lower than the value we derived, and thus leads to a lower mass that is in conflict with that from other methods. However, we believe that the value quoted in their paper is in error. When we reanalyze the 47 redshifts as given in their Table 1, we obtain a velocity dispersion of $1098\pm 157$ km s$^{-1}$. Therefore, there is actually no discrepancy between the dynamical mass estimate and that from X-ray and lensing analysis.

\subsubsection{Merger in the centre?}\label{mer}
\begin{figure*}
\includegraphics[width=9cm]{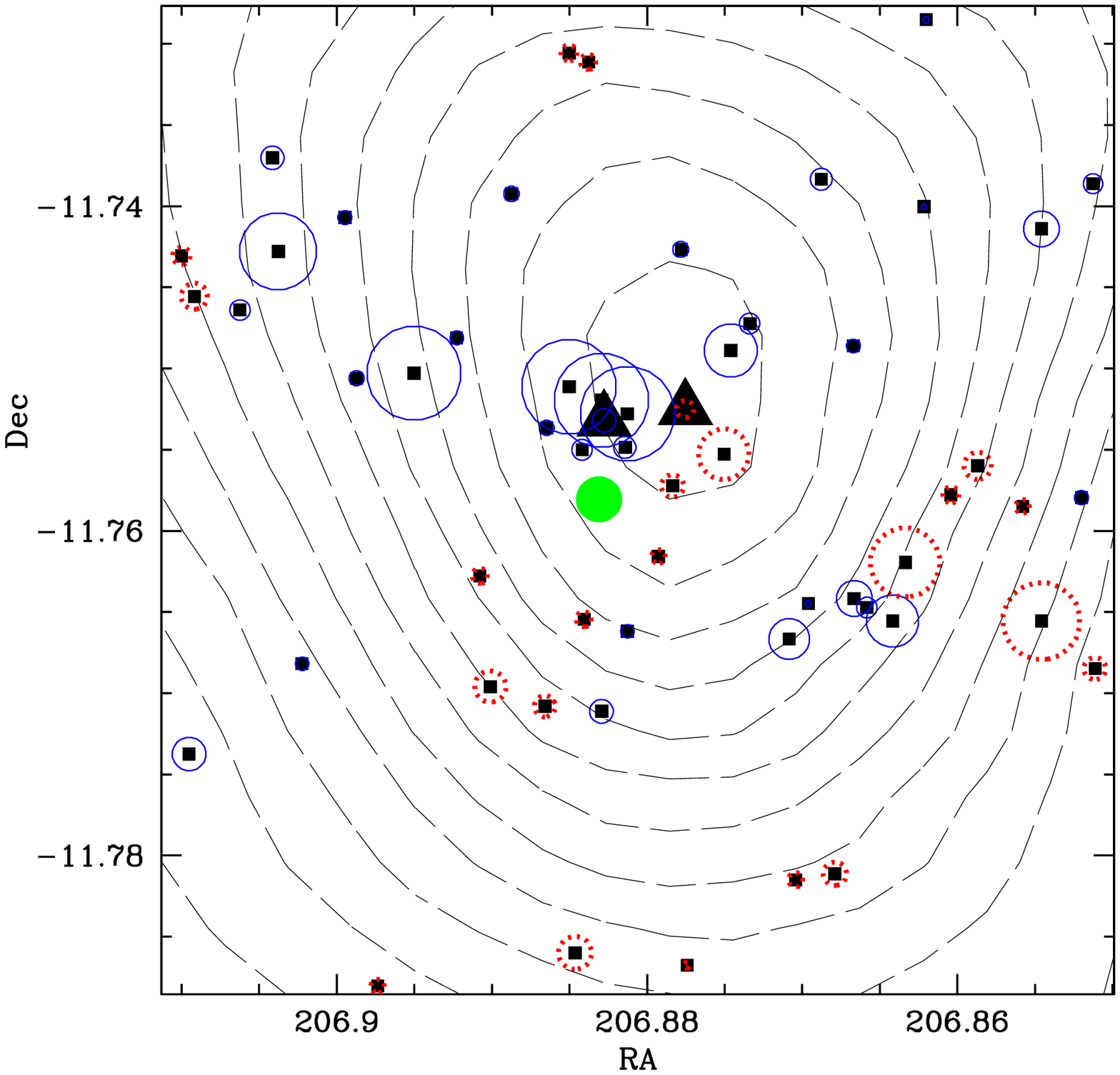}\includegraphics[width=9cm]{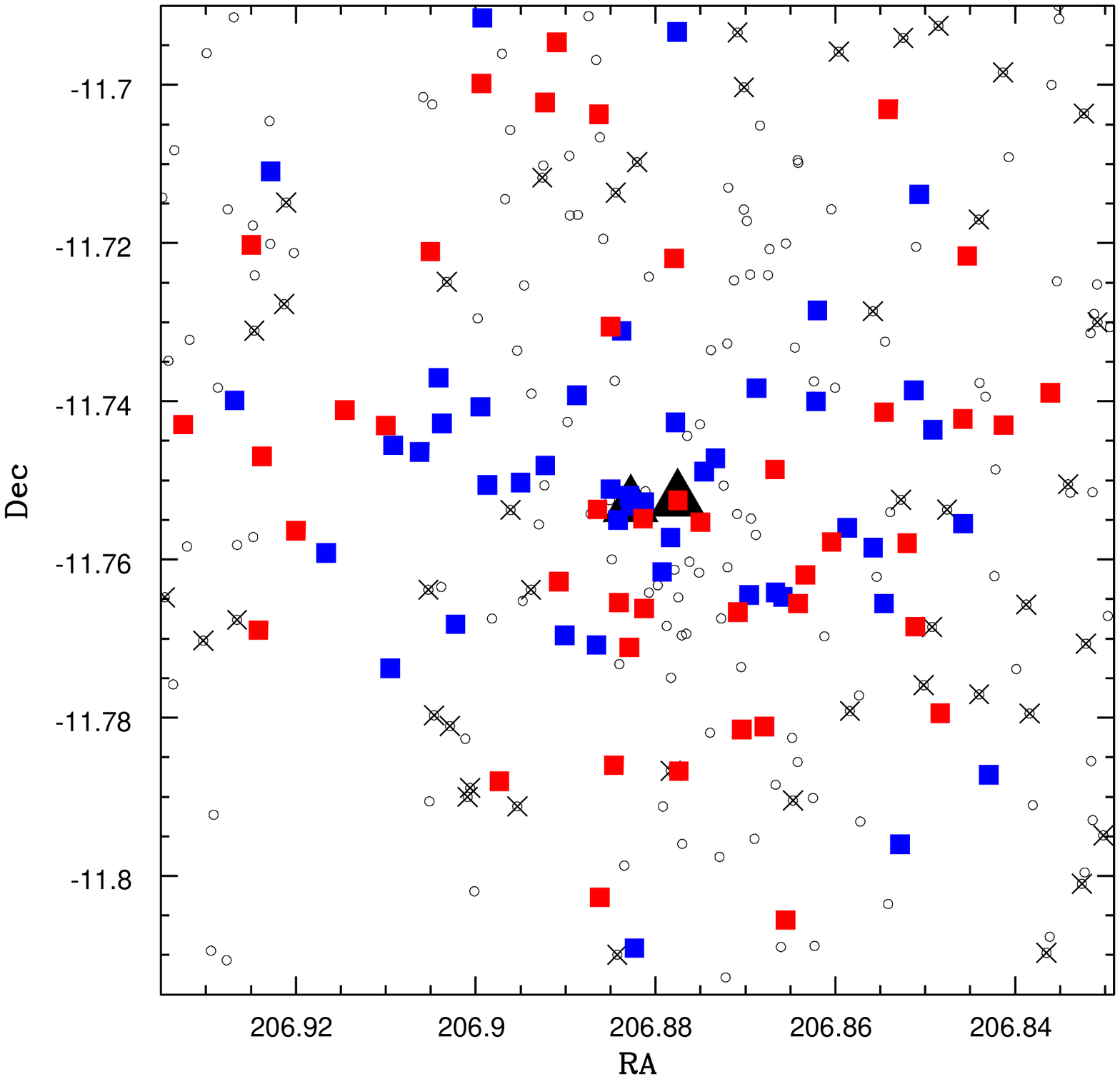}
\caption{Left panel: the spatial distribution of galaxies in the centre of the main cluster (the field of view is $\sim$1 Mpc $\times$ 1 Mpc), with the contours of the weak lensing signal-to-noise mass map overlaid, with an interval of 0.5$\sigma$. For clarity, we only plot the confirmed cluster members (the solid squares). The two triangles indicate the two BCGs. The size of the open circles  (proportional to $\delta$) around the cluster members show the local deviation from the global mean, with solid open circles indicating negative local deviations ($\bar {z}<0.4513$) and dashed open circles positive local deviations ($\bar {z}>0.4513$). The big filled green circle indicates the region $\sim 20$ arcsec SE from cluster centre, where the shocked gas is detected.  See text for details. Right panel: the spatial distribution of galaxies in the centre of the cluster, with a field of view of $\sim$2 Mpc $\times$ 2 Mpc. The open circles indicate possible potential members (not targeted), the crosses indicate nonmembers, the blue solid squares represent confirmed members with $z<0.4513$, and the red solid squares represent confirmed members with $z>0.4513$. Again the two triangles indicate the two BCGs.\label{dres}}
\end{figure*}
One important reason the merger scenario was proposed for this cluster is to explain the large discrepancy between the mass estimated from the velocity dispersion by \cite{ck} and the X-ray and lensing mass estimates discussed in the above section. However, as shown above, our new measurement resolves this discrepancy, and thus removes the original motivation for believing the system is undergoing a major merger.

Nonetheless, there is other evidence suggesting that this cluster might have recently undergone a merger. Although the X-ray morphology is generally relaxed in the centre of the cluster, the existence of a region with shocked gas  $\sim 10-$20 arcsec ($\sim 0.09$ Mpc) SE of the cluster centre has been detected as a high-pressure region in X-ray observations \citep{allen02,gitti04,ota08,bradac08,mir08} and a strong decrement in the Sunyaev-Zel'dovich observations \citep{komatsu01}. Also, \cite{mir08} found that their strong lensing data can be much better described by a two component model with comparable mass than the single mass component model. The fact that there are two very bright galaxies in the centre of this cluster 18$^{\prime \prime}$ ($\sim$ 0.1 Mpc) apart in the plane of sky is also taken as a possible indication of a merger event (but note their line-of-sight relative velocity is only $\lsim $100 km/s). \cite{kita} suggested that one possible scenario is a collision of equal mass clumps with a collision speed of 4600 km s$^{-1}$. However, this collision cannot be mostly along the line-of-sight, because as shown in \S \ref{vdisp} (and also in the work by \citealt{ck}), the redshift distribution is better described as a single Gaussian distribution. As discussed in \S \ref{vdisp}, our velocity data are also consistent (at the 30\% significance level), with a model distribution having two Gaussian components. However, even in that case, the radial velocity offset between the two peaks is only about 1000 km s$^{-1}$, much smaller than 4600 km s$^{-1}$. Furthermore, if the centre of the cluster is indeed composed of two merging components, each with a much smaller velocity dispersion than what we find for the system as a whole (see \S \ref{vdisp}), the dynamically--estimated mass profile would be much lower than that shown in Figure \ref{mcau} (the hexagon and the filled points), and  that would be inconsistent with the  weak lensing and X-ray mass estimates.

To investigate this further, we use the $\Delta$-test proposed by \cite{dr-sh} to examine closely the 3D substructures in the centre of the main cluster. This test measures the deviation of the local velocity and velocity dispersion from the global mean, and is quantified as :
\begin{eqnarray}
\delta^2=\frac{N_{nb}+1}{\sigma^2}\left [  (\bar{v}_{local}-\bar{v})^2+ ({\sigma}_{local}-{\sigma})^2   \right ],
\end{eqnarray}
where $N_{nb}$ is the number of neighbour galaxies, $\bar{v}_{local}$ and ${\sigma}_{local}$ are local mean velocity and dispersion measured from each galaxy and its $N_{nb}$ nearest neighbours, and $\bar{v}$ and $\sigma$ are global mean velocity and velocity dispersion, respectively.

 In the left panel of Figure \ref{dres}, we show the spatial distribution of the confirmed cluster members (solid squares) with their local deviation from the global mean indicated by the size (proportional to $\delta$) of the circle around them. The solid circles indicate negative local deviations ($\bar {z}<0.4513$), and dashed open circles positive local deviations ($\bar {z}>0.4513$). To examine closely the centre of the cluster, in this panel we only plot the region within a radius of $\sim 0.5$ Mpc from the cluster centre. The two triangles indicate the two BCGs. The region $\sim 20$ arcsec SE of cluster centre, indicated by the large green filled circle, is where the shocked gas is detected. Around that region, we do not detect significant deviation from the global mean. However, there seems to be a hint of a segregation between the region NE and SW of the cluster centre, in the sense that in the NE region most galaxies have negative local mean velocity relative to the mean redshift of the cluster, while galaxies in the SW region mostly have positive local mean relative velocity. This segregation is along the same direction connecting the main cluster and RXJ1347-SW. 

Note the above estimated local mean might be biased by a few galaxies with large deviations due to our sparse data sampling. Therefore, in the right panel of Figure \ref{dres}, we examine directly the spatial distribution of the redshift of individual galaxies, with a field of view of $\sim$2 Mpc $\times$ 2 Mpc. The open circles indicate possible potential members (not targeted), and the crosses represent nonmembers. Cluster members with $z<0.4513$ are indicated by the blue solid squares, and members with  $z>0.4513$ are indicated by the red solid squares. There is  a hint that more galaxies with negative relative velocity reside in the region NE to the cluster centre. To test whether the distribution of the two samples is significantly different, we apply a 2-dimensional Kolmogorov-Smirnov test following the procedure proposed by \cite{2dks}. The test suggests that there is a $\sim$60 per cent probability that galaxies with negative and positive relative velocity have different spatial distributions.   

The above evidence discussed in this section suggests that this might be a merger more in the plane of the sky than along the line-of-sight. However, with the lack of both a clear sharp shock front and a distinct separation of mass concentration on both sides of the shock (see contours in the left panel of Figure \ref{dres}), as observed in the famous example of merger in the plane of the sky, the Bullet cluster \citep[e.g.][]{bullet,clowe06}, the exact merger scenario still needs closer examination both in observations and simulations.

\subsection{A possible filamentary connection}\label{simu}
As seen in Figure \ref{vr}, cluster RXJ1347-SW  is $\sim 7$ Mpc (projected in the plane of sky) away from the main cluster, and has a rest-frame velocity offset of $\sim$4000 km s$^{-1}$ relative to the main cluster. This could be a result of two scenarios. One is that RXJ1347-SW is at nearly the same distance as the main cluster, and is falling into the cluster at a speed of  $\sim 4000$ km s$^{-1}$. The other is that it is $\sim 25$ Mpc in the background moving with Hubble expansion. We resort to simulations to try distinguishing these two scenarios. Since what we are looking for is how likely it is for two clusters like RX J1347-1145 and RXJ1347-SW to have a relative peculiar velocity as large as observed, we use the Hubble volume simulations \citep{hubble} because of its large box size ($3000^3 h^{-3} $Mpc$^{3}$).

From the cluster catalogue in the full-sky light cone with $\Lambda$CDM cosmology ($\Omega=0.3, \Lambda=0.7, \sigma_8=0.9$), we search, in a redshift range close to the redshift of RXJ1347, for a pair of clusters that have similar masses ($M1 \gsim 1\times 10^{15} M_{\odot}, M2 \gsim 1\times 10^{14} M_{\odot}$), projected distance ($\sim 10$ Mpc), and rest-frame velocity offset as RX J1347-1145 and RXJ1347-SW. We do not find any system where they are at the same distance, but with an observed velocity offset of $\sim 4000$ km s$^{-1}$.  This does not mean definitely that no system could have such a large relative peculiar velocity, because this is based on only one simulation with a large but still limited volume, under the assumption of one particular set of cosmological parameters. Among these parameters, $\sigma_8$ is the most uncertain one, and in fact a lower value than used in the simulation is preferred by the latest WMAP measurement \citep{wmap5}. However, in that case, the peculiar velocity between clusters would be even lower due to the lower attracting force from the lower mass density. Even though we cannot reach a definitive conclusion, this does suggest that the probability that RXJ1347-SW is falling into the main cluster with such a high velocity is low; instead, it is more likely to be in the background. 

\begin{figure}
\leavevmode \epsfysize=8cm \epsfbox{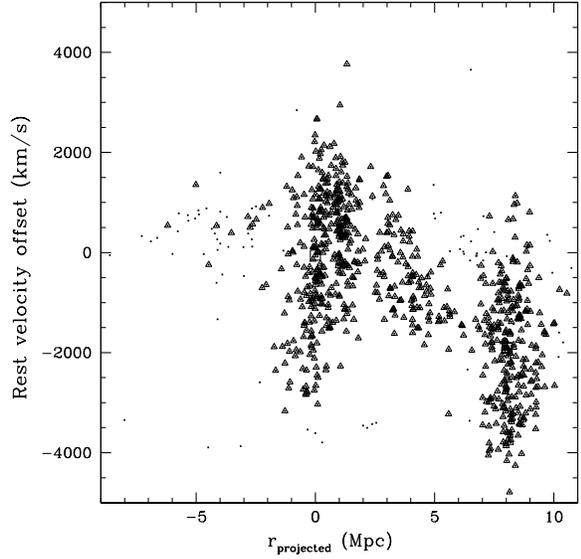}
\caption{This figure shows the rest-frame velocity offset of galaxies around two simulated clusters, as a function of distance projected in the direction connecting the two. The two clusters ($M_1=2.7\times 10^{15} M_{\odot}$, $M_2=1.8\times 10^{15} M_{\odot}$) are connected by a real physical filament traced by galaxies, with two other clusters in between. The excess of galaxies between them in velocity space is clearly visible. The two squares indicate the centres of the two clusters, and the triangles indicate galaxies that are spatially within a cylinder with a radius of 3 Mpc connecting the two clusters. \label{simucon}}
\end{figure}

However, just because RXJ1347-SW is far in the background does not necessarily mean
that there is no connection between it and the main cluster. The fact
that we observed an excess of galaxies between the two in velocity
space (Figure \ref{vr}) suggests there might be a physical connection
(see \citealt{ta09} for a report of the spectroscopic
detection of a large-scale galaxy filament associated with a massive
cluster at $z=0.5$;  also \citealt{ebeling04}). We compare our observation with simulations to
see if the observed excess of galaxies could represent a real
large-scale filamentary structure connecting the main cluster and
RXJ1347-SW, rather than foreground and background galaxies {\it projected}
in velocity space. To examine this on the scale of individual
galaxies, we make use of the galaxy catalogue from a semi-analytic
model by \cite{font}\footnote{Note for our purpose here the exact
  semi-analytic model used is unimportant.} based on the high
resolution Millennium simulation \citep{ms}.  Because of the small
volume of the simulation, there are only four clusters that have mass
greater than $1\times 10^{15} M_{\odot}$. We search around these four
clusters for a second cluster with lower mass and at a similar
distance projected in the plane of sky and along the line-of-sight
from the more massive cluster as in the case of our observed RX
J1347-1145 and RXJ1347-SW. For the four pairs of systems we found, we
calculate the line-of-sight velocity offset of each galaxy relative to
the more massive cluster, and plot it as a function of distance,
projected along the direction towards the less massive one. For direct
comparison with our observed data, we only plot galaxies in a strip
with a width similar to the radius of the FOV as our spectroscopic
observation, down to the same magnitude limit as our data. In Figure
\ref{simucon} we show one extreme case, where the two clusters
($M_1\sim 3\times 10^{15} M_{\odot}$, $M_2\sim 2\times 10^{15}
M_{\odot}$) are connected by a physical, large-scale filament traced by
galaxies, with two other clusters in between. This structure is
clearly visible in velocity space. The triangles indicate galaxies
that are spatially within a cylinder with a radius of 3 Mpc
(comparable to the virial radius of the more massive cluster)
connecting the two clusters. This is an extreme case where the galaxy
filament is very impressive due to the presence of two other massive
systems in between the two clusters. However, even in other cases in
the simulation where the excess of galaxies between two clusters is
smaller ($\sim 5-10$) than this one, the excess is still due to
galaxies that {\it physically reside} in the region between two clusters,
instead of being purely due to a {\it projection} effect. Although this is not definitive, it does support the possibility that 
the excess of galaxies we observe in velocity space between the two
clusters reflects a real physical structure.

\section{Summary}\label{sum}
In this work, we presented photometric and spectroscopic analysis on the dynamics and large-scale structure around the most X-ray luminous cluster RX J1347-1145, and found the following results:\\
$\bullet$ We identified an elongated large-scale structure around the main cluster, photometrically using red-sequence galaxies, and discovered a rich cluster (RXJ1347-SW) $\sim 7$ Mpc (projected) away from the main cluster.\\
$\bullet$ The velocity dispersion of the main cluster is 1163$\pm 97$ km s$^{-1}$, corresponding to  $M_{200}= 1.16^{+0.32}_{-0.27} \times 10^{15} M_{\odot}$, with  $r_{200}=1.85$ Mpc.\\
$\bullet$ Our dynamical mass estimates based on both the velocity dispersion and velocity caustics are consistent with our weak lensing analysis, and the X-ray studies in the literature. We also find excellent agreement between these mass estimates and that estimated photometrically from the abundance of red-sequence galaxies (the richness). \\
$\bullet$ Our new measurement resolves a previous claimed discrepancy between the dynamical mass estimate and other methods \cite{ck}.\\
$\bullet$ Our spectroscopic analysis confirmed that the photometrically identified structure  RXJ1347-SW is indeed a massive cluster, with a velocity dispersion of $780\pm 100$ km s$^{-1}$. \\
$\bullet$ Our weak lensing analysis shows good correspondence between the mass concentration and the overdensity of red-sequence galaxies in the main cluster and RXJ1347-SW.\\
$\bullet$ Comparing with simulations, the observed excess of galaxies between the main cluster and RXJ1347-SW relative to other regions around the main cluster in velocity space suggests a possible filamentary structure connecting these two clusters. This structure has a large population of blue galaxies associated with the overdensity of red-sequence galaxies, suggesting that targeting overdensities of red-sequence galaxies for spectroscopic follow-up  allows the efficient selection of cluster members, while still tracing interesting structures with active star formation.\\

This work is a first result from our large on-going spectroscopic campaign. In future work, we will use the technique presented in this work to search and examine the properties of other structures, especially groups in the infall region, with the on-going data collecting.

\section*{Acknowledgment}

This work was supported by an Early Researcher Award from the province of Ontario and an NSERC Discovery Grant to MLB. We thank Renbin Yan for providing the pipeline for redshift determination, and Eiichiro Komatsu for useful discussion, and Magellan telescope operators for the help during the observation. TL thanks Sean McGee for helpful discussions.

The simulations in this paper were carried out by the Virgo Supercomputing Consortium using computer s based at the Computing Centre of the Max-Planck Society in Garching and at the Edinburgh parallel Computing Centre. The data are publicly available at http://www.mpa-garching.mpg.de/NumCos.

\appendix
\section{Redshift Completeness, Success Rate and Efficiency}\label{cp}

\begin{figure}
\leavevmode \epsfysize=8cm \epsfbox{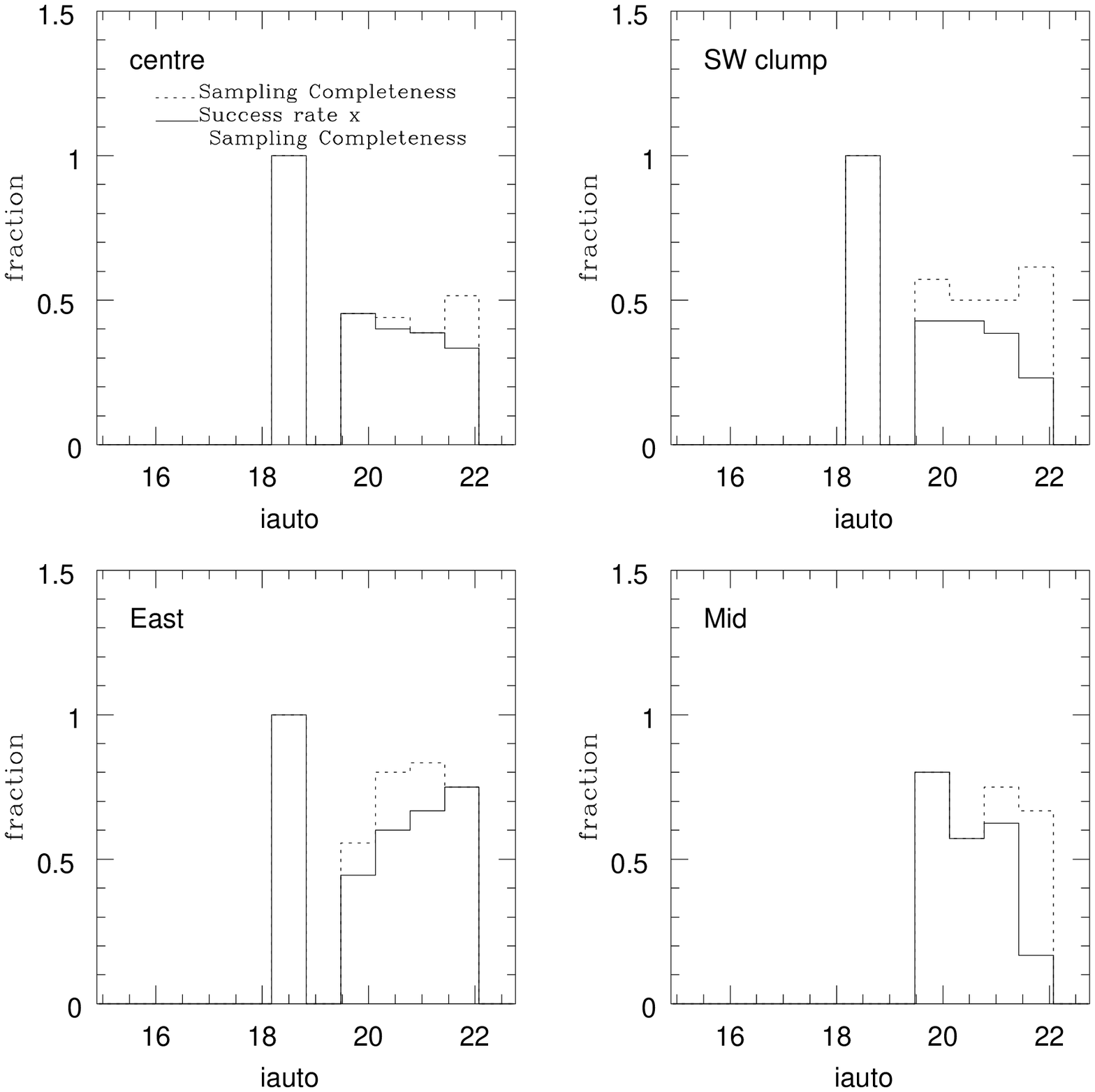}
\caption{The completeness of spectroscopic follow-up and success rate for red galaxies in the 4 pointings. The dashed lines show the sampling completeness (number of galaxies targeted / number of all targets), and the solid lines show the overall completeness, i.e. the success rate (number of galaxies with secure redshift obtained / number targeted) multiplied by the sampling completeness.
\label{cpred}}
\end{figure}

\begin{figure}
\leavevmode \epsfysize=8cm \epsfbox{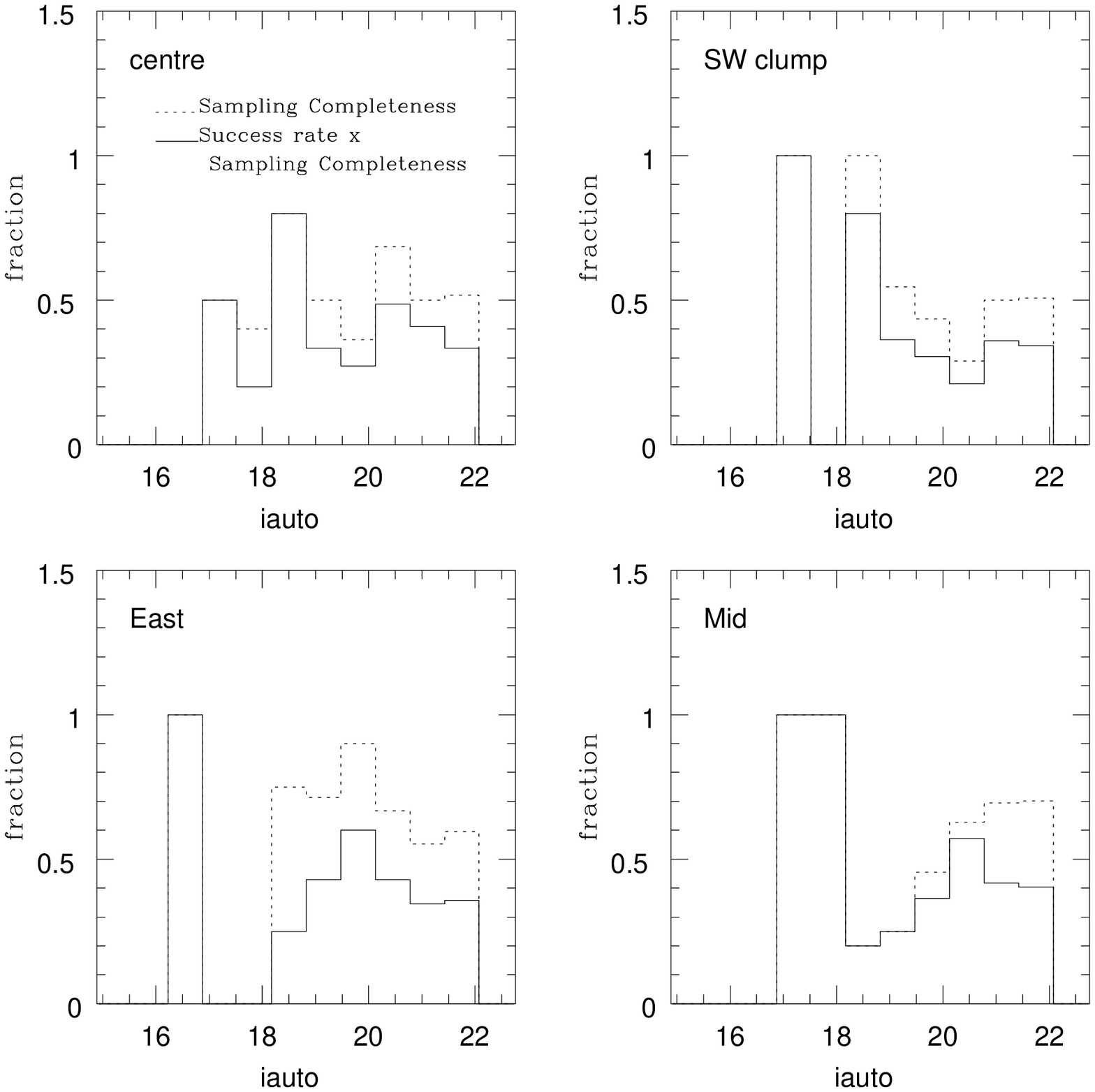}
\caption{The same as Figure \protect \ref{cpred}, but for the blue population.
\label{cpblue}}
\end{figure}

In Figure \ref{cpred} and \ref{cpblue}, we show the sampling completeness (number of galaxies targeted / number of all targets), and overall completeness, i.e. the success rate (number of galaxies with secure redshift obtained / number targeted) multiplied by the sampling completeness, for the red and blue population in the 4 pointings, respectively. Our overall completeness is about 50 per cent.

 For red galaxies, in pointings ''Centre'' and ''SW'', about 90 per cent of galaxies with secure redshifts obtained turned out to be members; and in the rest of the two less densed pontings, the efficiency is $\sim$60 per cent and $\sim$70 per cent. For blue galaxies, the typical efficiency is about 30 per cent. (The relatively low efficiency of obtaining blue members is not surprising.)


\section{Redshift Uncertainty}\label{zun}

\begin{figure}
\leavevmode \epsfysize=8cm \epsfbox{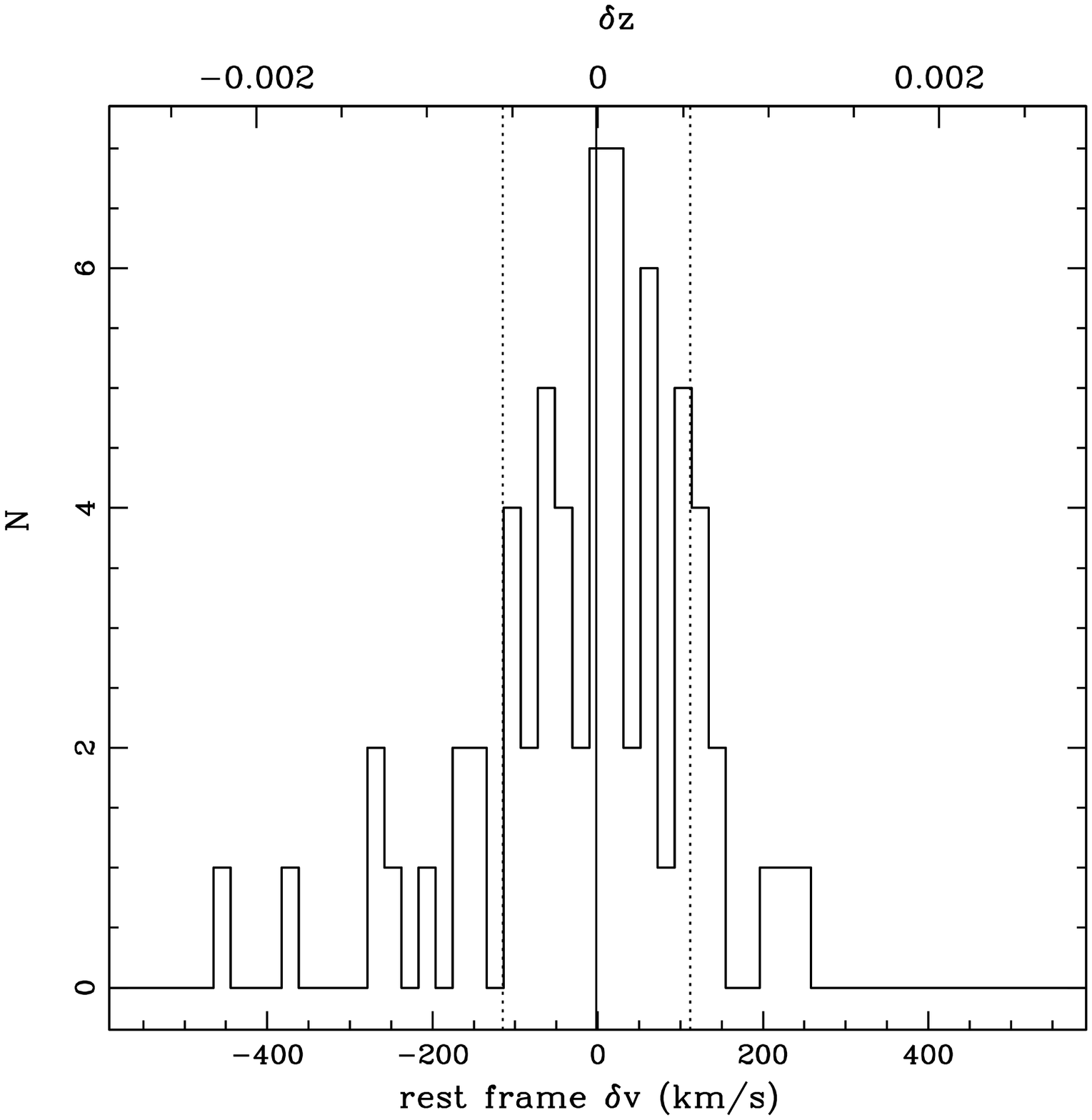}
\caption{The distribution of the difference in redshift (top axis) or rest-frame velocity (bottom axis) for galaxies with repeated observations in our sample. The vertical solid line is the mean of the distribution, and the vertical dashed lines indicate 1$\sigma$ dispersion of the distribution.
\label{dup}}
\end{figure}

We possess duplicate observations for some of the objects that appear on both masks at each pointing. To determine the uncertainty on the redshift, we compare the difference in the redshifts of the same object from two observations in Figure  \ref{dup}. The histogram is the distribution of the difference in redshift (top axis) or rest-frame velocity (bottom axis) from repeated observations. The vertical solid line is the mean of the distribution, and vertical dashed lines indicate 1$\sigma$. It shows that the uncertainty on our redshift is $\sim$113 km s$^{-1}$ in the rest frame.

\begin{figure}
\includegraphics[width=45mm]{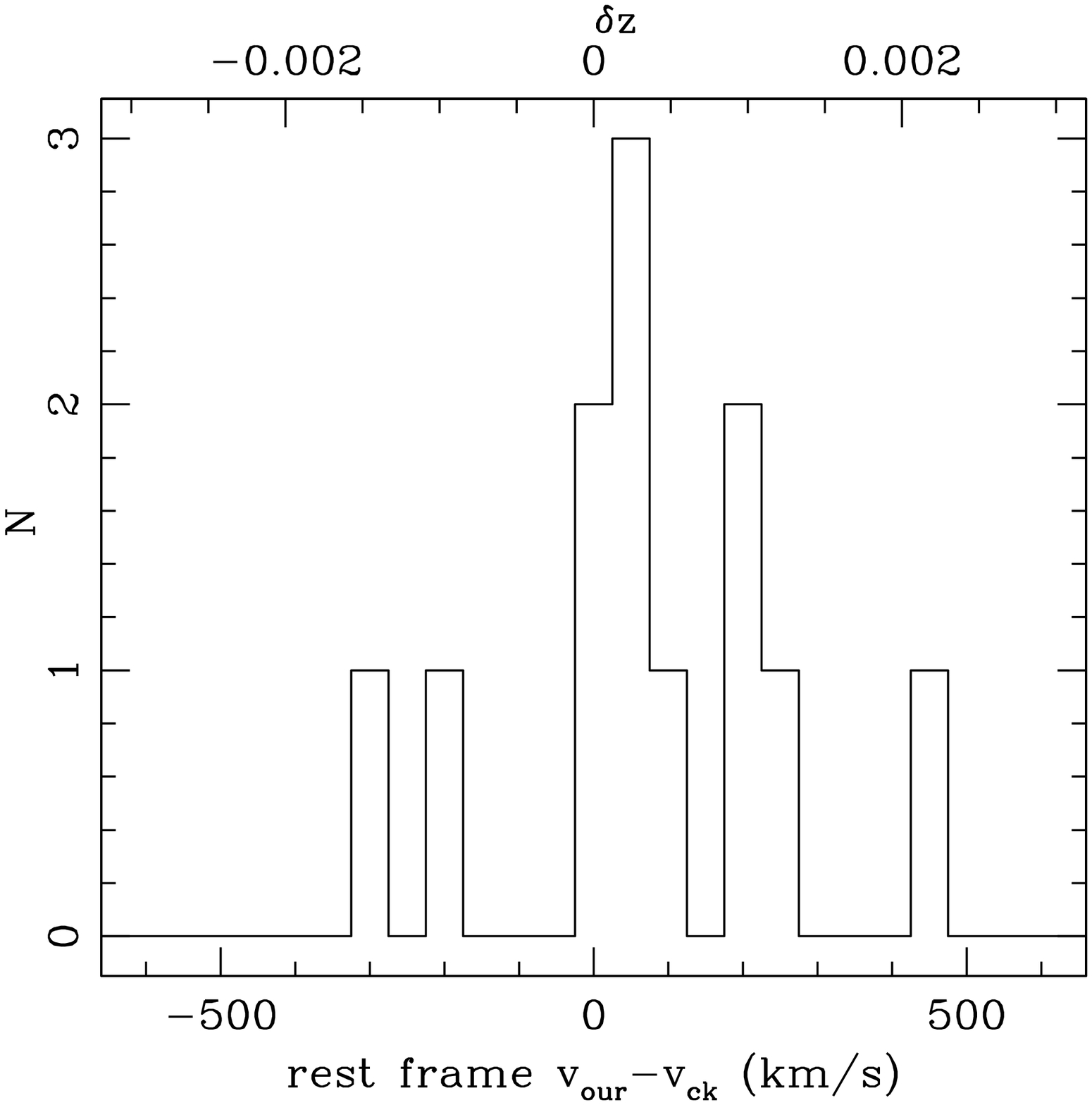}\includegraphics[width=43mm]{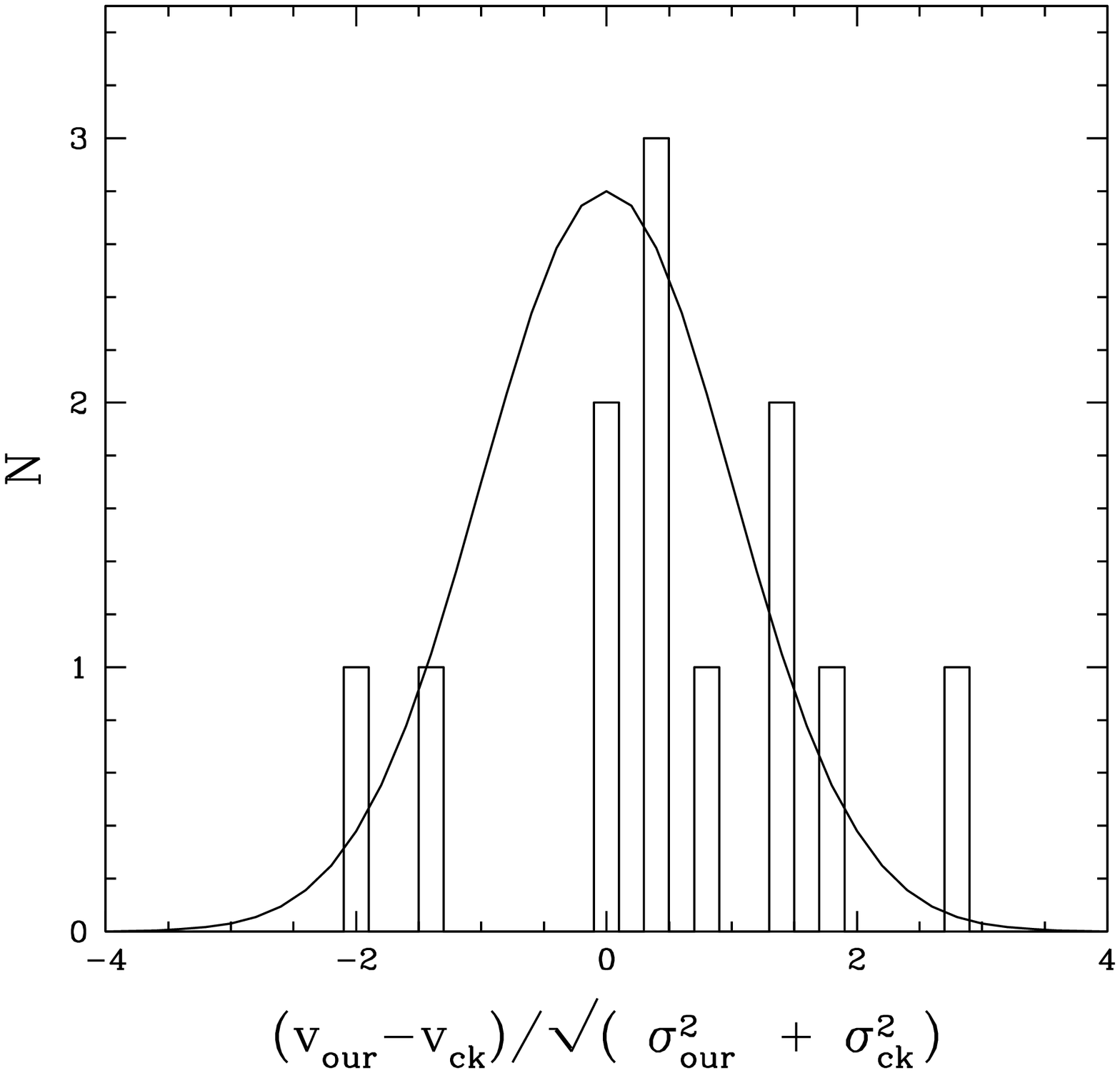}
 \caption{ Left panel: the distribution of the difference in redshift (top axis) and rest-frame velocity (bottom axis) for common objects between our sample and the sample of \protect \cite{ck}.  Right panel:  the distribution of the difference in  the rest-frame velocity  for the common objects normalized by the quadratically added uncertainties on our redshifts and those of \protect \cite{ck}. The curve is a standard normal distribution. \label{cohenz}}
\end{figure}

We also compare the redshifts determined from our observation with that from \cite{ck} for the matched objects. The distribution of the difference is shown in the left panel of Figure \ref{cohenz}, with top axis in unit of redshift and bottom axis rest-frame velocity. In the right panel of  Figure \ref{cohenz}, the distribution of the difference in terms of the rest-frame velocity is normalized by the quadratically added uncertainties on our redshift (113 km s$^{-1}$) and that of \cite{ck} (which they assumed to be 100 km s$^{-1}$). This distribution is not significantly different from a standard normal distribution shown as the curve, although it is a small number statistics. This suggests that there is no systematic offset between the redshifts measured by \cite{ck} and us; therefore, we add the extra 34 redshift from \cite{ck} to our sample when applicable.

\section{Sample Spectra}\label{eg}
\begin{figure}
\includegraphics[width=87mm]{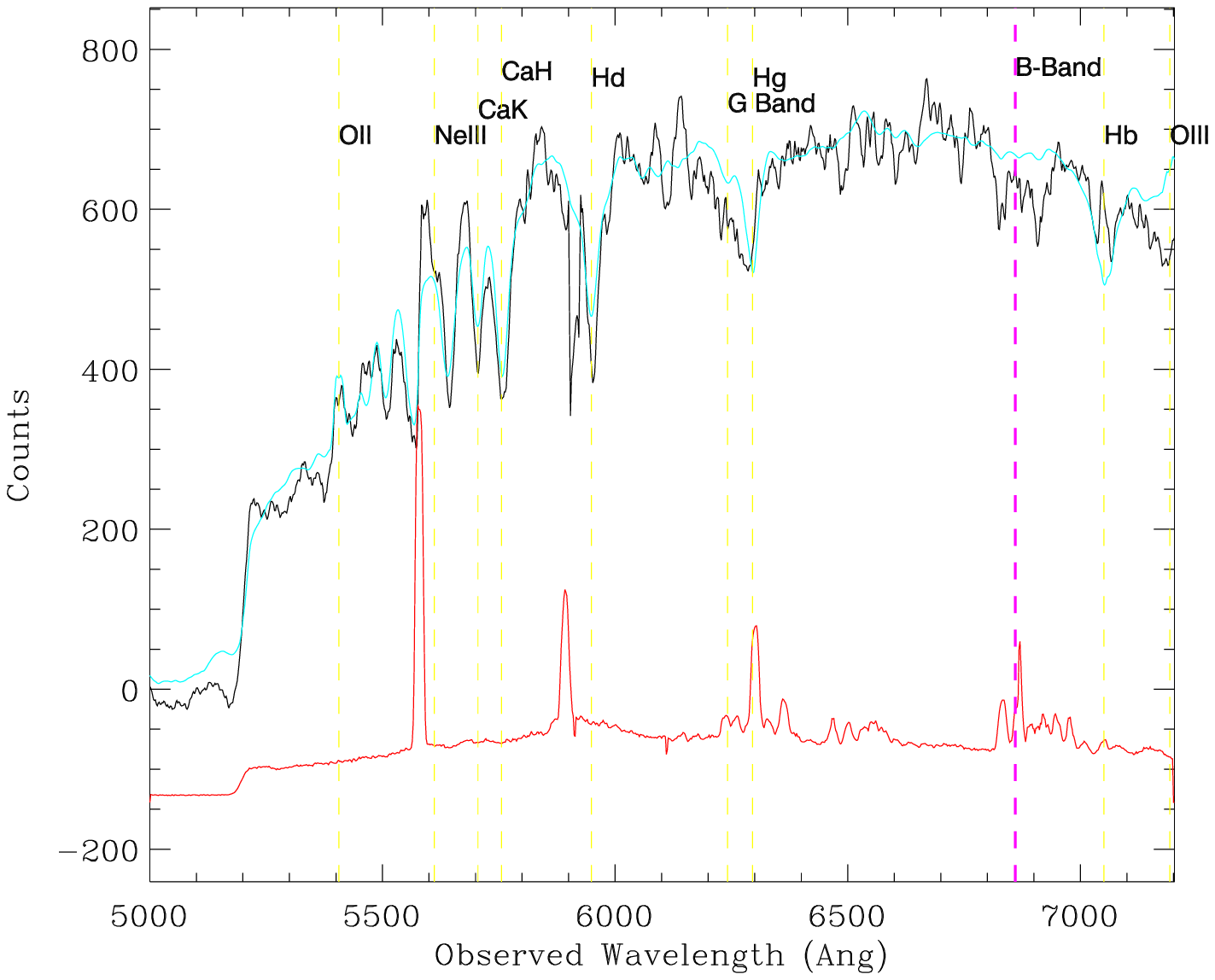}
\includegraphics[width=87mm]{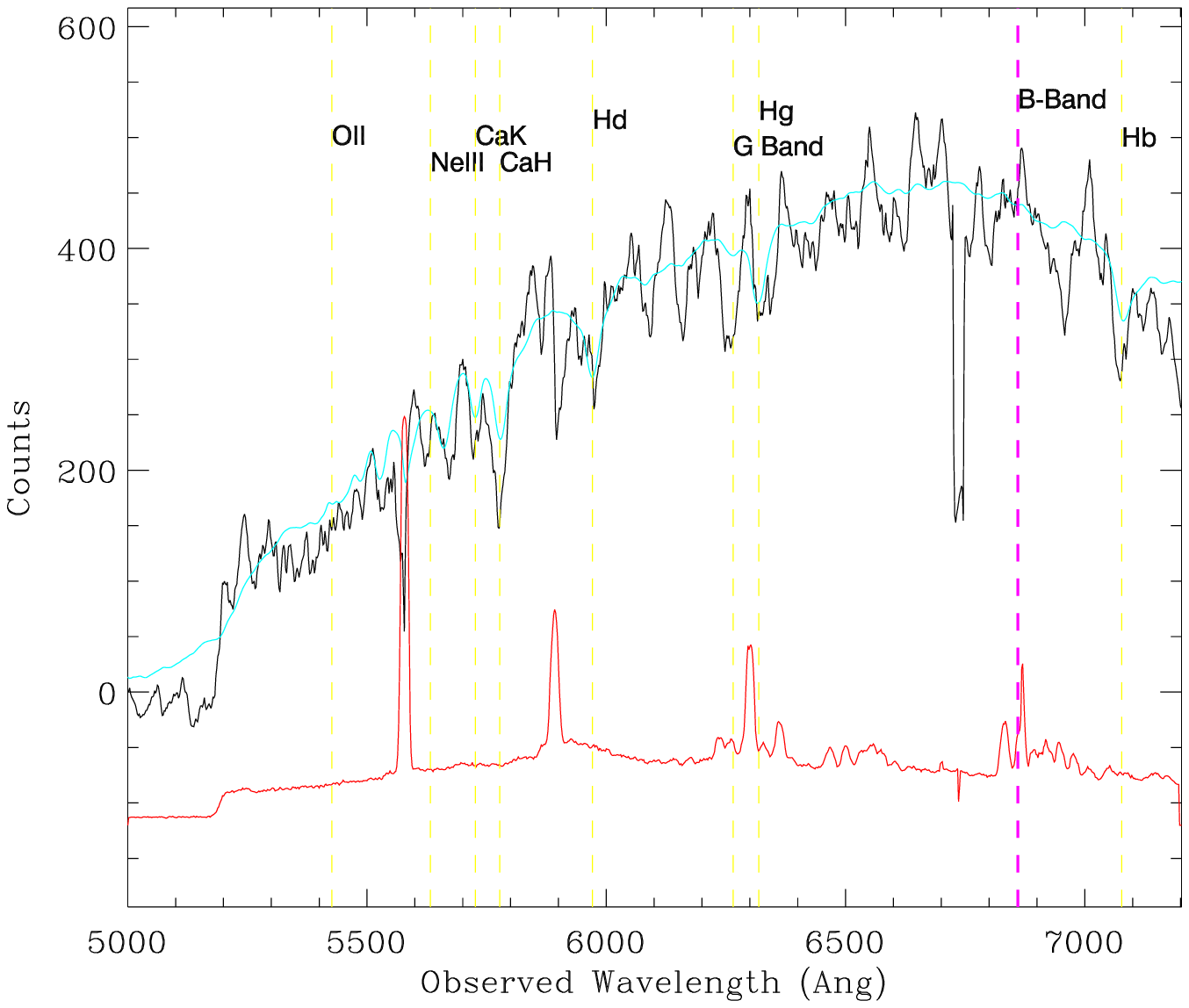}
\caption{Example spectra (the black line) of two typical galaxies with $i'$=21.4 mag and $i'$=21.8 mag, close to the magnitude limit of our sample. The cyan line shows the template spectrum. The red line shows the sky spectrum, arbitrarily shifted and renormalized, to indicate where bright sky lines give rise to residuals.  The positions of different spectral features are indicated with vertical lines, and in both spectra there are multiple absorption features identified, ensuring a secure redshift.
\label{egspec}}
\end{figure}

Figure \ref{egspec} shows example spectra of two typical absorption-line galaxies with $i'$=21.4 mag and $i'$=21.8 mag, close to the magnitude limit of our sample. Multiple absorption features are visible, including prominent CaK, CaH, H$\delta$, H$\beta$ lines. See figure caption for description of different lines.

\end{document}